\pdfoutput=1

 \documentclass{emulateapj}

\usepackage{graphics}

\newcommand {\Lya}    {Ly$\alpha$}   

\newcommand {\HI}        {\ion{H}{1}}   

\newcommand {\HeI}     {\ion{He}{1}}   
\newcommand {\HeII}     {\ion{He}{2}}   

\newcommand {\OIV}    {\ion{O}{4}}   
\newcommand {\OV}    {\ion{O}{5}}    
\newcommand {\OVI}    {\ion{O}{6}}   
\newcommand {\OVII}   {\ion{O}{7}}
\newcommand {\OVIII}  {\ion{O}{8}}

\newcommand {\CIV}    {\ion{C}{4}}
\newcommand {\CV}     {\ion{C}{5}}
\newcommand {\CVI}    {\ion{C}{6}}

\newcommand {\NV}     {\ion{N}{5}}
\newcommand {\NVI}     {\ion{N}{6}}
\newcommand {\NVII}    {\ion{N}{7}}

\newcommand {\NeVIII}  {\ion{Ne}{8}}   
\newcommand {\NeIX}  {\ion{Ne}{9}}

\newcommand {\NHI}    {$N_{\rm HI}$}
\newcommand {\NOVI}   {$N_{\rm OVI}$}

\newcommand {\kms}    {km~s$^{-1}$}

\newcommand {\cd}     {cm$^{-2}$}  
\newcommand {\FUSE}   {{\it FUSE}} 
\newcommand {\HST}    {{\it HST}}

\newcommand {\etal}   {et~al.}

\begin{document}

\title{The Baryon Census in a Multiphase Intergalactic Medium: \\
30\%  of the Baryons May Still Be Missing}  

\author{J. Michael Shull, Britton D. Smith\altaffilmark{1},  \& Charles W. Danforth }
\affil{CASA, Department of Astrophysical \& Planetary Sciences, \\
University of Colorado, Boulder, CO 80309}

\altaffiltext{1}{now at Department of Physics and Astronomy, Michigan State University, East Lansing, MI 48824} 

\email{michael.shull@colorado.edu,  charles.danforth@colorado.edu, smit1685@msu.edu}  


\begin{abstract} 

Although galaxies, groups, and clusters contain $\sim10$\% of the baryons, many more reside in the photoionized 
and shocked-heated intergalactic medium (IGM) and in the circumgalactic medium (CGM).  We update the baryon 
census in the (\HI)  \Lya\ forest and warm-hot IGM (WHIM) at $10^{5-6}$~K traced by \OVI\ $\lambda1032,1038$ 
absorption.  Using robust cosmological simulations of heating, cooling, and metal transport, we improve the \OVI\ 
baryon surveys with spatially averaged corrections for metallicity ($Z / Z_{\odot}$) and \OVI\ ionization fraction 
($f_{\rm OVI}$).  Statistically, their product correlates with column density, 
$(Z / Z_{\odot}) \, f_{\rm OVI}\ \approx (0.015)(N_{\rm OVI} /10^{14}\,{\rm cm}^{-2})^{0.70}$, with a
\NOVI-weighted mean of 0.01, which doubles previous estimates of WHIM baryon content.  We also update
the \Lya\ forest contribution to baryon density out to $z = 0.4$, correcting for the $(1+z)^3$ increase in absorber 
density,  the $(1+z)^{4.4}$ rise in photoionizing background, and cosmological proper length $d\ell/dz$. 
We find substantial baryon fractions in the photoionized \Lya\ forest  ($28 \pm 11$\%) and WHIM traced by \OVI\ and 
broad-\Lya\ absorbers $(25 \pm 8$\%).  The collapsed phase (galaxies, groups, clusters, CGM) contains 
$18\pm4$\%, leaving an apparent baryon shortfall of $29\pm13$\%.  Our simulations suggest that $\sim15$\%
reside in hotter WHIM ($T \geq 10^6$~K).  Additional baryons could be detected in weaker \Lya\ and \OVI\ absorbers.   
Further progress requires higher-precision baryon surveys of weak absorbers, down to minimum column densities 
$N_{\rm HI} \geq  10^{12.0}$~\cd,  $N_{\rm OVI} \geq  10^{12.5}$~\cd, $N_{\rm OVII} \geq 10^{14.5}$~\cd, using
high-S/N data from high-resolution UV and X-ray spectrographs.  

\end{abstract} 



\section{INTRODUCTION}

For low-redshift cosmology and galaxy formation rates, it is important to account for all the baryons synthesized 
in the Big Bang.   Cosmologists have noted a baryon deficit  in the low-redshift universe (Fukugita, Hogan, \& Peebles 
1998) relative to the predicted density synthesized in the Big Bang.  Although this deficit could arise from an incomplete 
inventory, it could also challenge our understanding of the thermodynamics of structure formation and the response of
the gas to accretion shocks and galactic outflows.  Recent analysis (Komatsu \etal\ 2011) of the spectrum of acoustic 
peaks in the Cosmic Microwave Background (CMB) obtained by the {\it Wilkinson Microwave Anisotropy Probe} (WMAP) 
found that baryons comprise a fraction $\Omega_b = 0.0455 \pm 0.0028$ of the critical matter-energy density of the
universe, $\rho_{\rm cr} = (9.205 \times 10^{-30}~{\rm g~cm}^{-3}) h_{70}^2$, where $h_{70}$ is the Hubble constant  
($H_0$) in units of 70 km~s$^{-1}$~Mpc$^{-1}$.  This 4.6\% baryon fraction corresponds to a mean comoving density, 
$\overline{\rho}_b = \Omega_b \rho_{\rm cr} = 4.24 \times 10^{-31}$ g~cm$^{-3}$, and a hydrogen number density 
$n_H = 1.90 \times 10^{-7}~{\rm cm}^{-3}$, assuming a primordial helium mass fraction $Y_p = 0.2477$ 
(Peimbert \etal\  2007).  

Baryon inventories in the low redshift universe are uncertain.  They are often complicated by the formation of galaxies 
and large-scale structure and by the feedback from star formation in the form of ionizing radiation, metals, and outflows.  
Galaxy surveys have found $\sim$10\% of these baryons in collapsed objects such as galaxies, groups, and clusters 
(Salucci \& Persic 1999; Bristow \& Phillipps 1994; Fukugita \& Peebles 2004).   Over the last 15 years, substantial reservoirs 
of gas have been found in the intergalactic medium (IGM), in the halos of galaxies, and in the circumgalactic medium (CGM).   
There are often semantic problems in defining the CGM,  as gas blown out of galaxies or material within the virial radius  
(Tumlinson \etal\ 2011; Prochaska \etal\ 2011).  Of the remaining 80-90\% of cosmological baryons, approximately half can 
be accounted for in the low-$z$ IGM (Shull 2003;  Bregman 2007;  Danforth \& Shull 2008;  Danforth 2009) including the
warm-hot IGM (or WHIM).   Ultraviolet spectroscopic surveys of \Lya\ and 
\OVI\ have identified substantial numbers of absorbers (Danforth \& Shull 2008;  Tripp \etal\ 2008;  Thom \& Chen 2008),
but claimed detections of hotter  in X-ray absorption by \OVII\  (Nicastro \etal\ 2005) remain controversial
(Kaastra \etal\ 2006; Yao \etal\ 2012).  Unfortunately, X-ray spectra still have not confirmed the potential large reservoir of 
baryons at $T > 10^6$~K, suggested by cosmological simulations.  

An inefficient distribution of collapsed baryons vs.\ distributed matter is a prediction of nearly all cosmological simulations 
of large-scale structure formation (Cen \& Ostriker 1999, 2006;  Dav\'e \etal\ 1999, 2001; Smith \etal\ 2011; Tepper-Garcia
\etal\ 2011).   These N-body hydrodynamical simulations suggest that 10--20\% of the baryons reside in collapsed objects 
and dense filaments, with the remaining 80\%  distributed over a wide range of phases in baryon overdensity 
($\Delta_b = \rho_b / \overline{\rho}_b)$ and temperature ($T$).   Thermodynamic considerations suggest that shock-heated 
WHIM at $z < 1$ is a natural consequence of gravitational instability in a dark-matter dominated universe.  This hot gas is
augmented by galactic-wind shocks and virialization in galaxy halos.  

In this paper, we improve the analysis of baryon content in both photoionized and shocked-heated IGM phases and
assess the accuracy of observational and theoretical estimates.  Because \Lya\ absorption surveys probe the neutral 
component of diffuse, photoionized filaments, assessing the baryon content requires a correction for the neutral fraction,
$f_{\rm HI}$.   We derive more accurate photoionization corrections based on the $(1+z)^{4.4}$ rise in the metagalactic 
ionizing background out to $z \approx 0.7$, as well as the $(1+z)^3$ increase in absorber density and changes in 
cosmological proper length, $d\ell/dz$.   We also use cosmological simulations to correct the \OVI\ surveys for ionization 
fraction ($f_{\rm OVI}$) and metallicity  $(Z / Z_{\odot})$.  Finally, we use our group's critically evaluated catalog
of \HI\ and \OVI\ absorption lines  (Tilton \etal\ 2012) from our  \HST\ Legacy Archive project on IGM data taken by the 
{\it Hubble Space Telescope} (\HST) and {\it Far Ultravolet Spectroscopic Explorer} (\FUSE).

To constrain the distribution of $Z$ and $f_{\rm OVI}$, we use cosmological hydrodynamic simulations of IGM heating, 
cooling, and metal transport to find the column-density weighted average for their product,  $(Z / Z_{\odot}) \, f_{\rm OVI}$.  
Whereas previous work assumed constant values, $Z/Z_{\odot} = 0.1$ and $f_{\rm OVI} =0.2$, our new simulations provide 
spatially averaged corrections for metallicity, \OVI\ ionization fraction, and covariance of  IGM parameters ($T$, $\rho$, $Z$).  
Our computed mean value $\langle (Z / Z_{\odot}) \, f_{\rm OVI} \rangle = 0.01$ is half the previously assumed value, thereby
doubling previous estimates of the baryon census traced by \OVI.   In Section 2 we describe the simulations and their results for 
the \OVI\ distribution in column density, gas temperature, baryon overdensity, metallicity, and ionization fraction.  In Section 3,  
we assess the corrections for ionization and metallicity, applied to the \OVI\ and \Lya\  absorbers, and we derive values of  
$\Omega_b$ from recent surveys of IGM and CGM phases.   Variations in these factors are produced by WHIM thermodynamics,  
gas temperature ($T$), baryon overdensity ($\Delta_b$), and \OVI\ column density ($N_{\rm OVI}$).   In Section 4 and two 
Appendices, we summarize the current baryon census with uncertainties on each component.

\section{COSMOLOGICAL SIMULATIONS OF THE WHIM}

The primary simulation analyzed in this work is run $50\_1024\_2$ of Smith \etal\ (2011) with ``distributed feedback"
and performed with the adaptive-mesh-refinement + N-body code \texttt{Enzo} (Bryan \& Norman 1997;  O'Shea \etal\ 2004).   
To check convergence and robustness of our results, we also looked at other runs, particularly
run $50\_1024\_1$ with ``local feedback".  
Post-processing of the simulations was carried out using the data analysis and visualization package
\texttt{yt}\footnotemark, documented by Turk \etal\ (2011).\footnotetext{http://yt.enzotools.org/}
The simulation has a box size of $50h^{-1}$ Mpc comoving, with 1024$^{3}$ grid cells and dark matter particles, 
giving it a dark-matter mass resolution of $7\times 10^{6}\,M_{\odot}$ and spatial cells of $50h^{-1}$ kpc.  
Radiative cooling is included by solving for the non-equilibrium chemistry and cooling of atomic H and He, 
coupled to tabulated metal cooling rates computed as a function of density, metallicity, temperature, electron 
fraction, and redshift in the presence of an ionizing metagalactic radiation background.  
To mimic the effects of reionization, we included a spatially uniform, but redshift-dependent radiation 
background given by Haardt \& Madau (2001).  The influence of the radiation background 
through photoheating and photoionization is included in the cooling of both the primordial and metal species.

Our \texttt{Enzo} unigrid simulations (Smith \etal\ 2011), post-processed in the current paper, are among the best
current work in describing the temperature, metallicity, and ionization state of the hot gas (WHIM) and photoionized 
gas.   Our results have been checked for convergence (Smith \etal\ 2011), and they are robust. 
We have tested the results from several runs with different box sizes, resolutions, and modes of feedback 
(local and distributed), finding similar results, as described later.  
Our use of unigrid (rather than adaptive-mesh-refinement) hydrodynamics was deliberate, to provide optimal 
hydrodynamic resolution of the IGM and WHIM.   It is well established that SPH (smooth particle hydrodynamic) 
codes provide poor hydrodynamic resolution of shocks and instabilities.  Agertz \etal\ (2007) noted that Eulerian 
grid-based methods are able to resolve and treat important  dynamical instabilities such as Kelvin-Helmholtz or 
Rayleigh-Taylor, whereas these processes are poorly resolved by SPH techniques.  Vazza \etal\ (2007) pointed
to the significantly different phase diagrams of shocked cells in grid codes compared to SPH, with sizable differences 
in the morphologies of accretion shocks between grid and SPH methods.  

We have employed a sophisticated treatment of metal cooling and heating (Smith, Sigurdsson, \& Abel 2008; 
Smith \etal\  2011).  The SPH models by Dav\'e \& Oppenheimer (2009) and Dav\'e \etal\ (2011) find IGM phases with
markedly different temperatures and metallicities, but we believe these results arise from their metal-injection
schemes and the lack of mixing.  None of the other major simulation groups (Cen \& Ostriker 2006; Cen \& Chisari 2011; 
Tepper-Garcia \etal\ 2011; Smith \etal\ 2011; Cen 2012) can reproduce their results on IGM temperature and ionization
state.     Our $1024^3$ simulations have better hydrodynamic resolution than all of the Tepper-Garcia \etal\ simulations 
($512^3$) and Dav\'e \& Oppenheimer models ($384^3$).   Our paper was the only one to test for convergence, and 
we concluded that our $1024^3$ models were robust.   
It has sometimes been suggested that the use of ``unigrid" {\it Enzo} calculations casts doubt on the validity of the
results for metallicity production and transport, compared to adaptive-mesh-refinement (AMR).  Just the opposite
is true.  In fact, AMR does not provide a better computation of the global star formation rate, as we examined
(Smith \etal\ 2011).    In these simulations, star formation seems to be dependent on the force resolution on the root grid.  
Stars do not form in the lower-resolution simulations because the halos do not collapse correctly at early times when
there is no refinement occurring in the box.  By the time refinement happens, it is  already too late;  the collapse of the 
halos has already been delayed, owing to the poor force resolution.  We have verified this by running simulations, with 
and without AMR, at the same resolution on the top grid.   We find almost the exact same star-formation history.  For this
reason, we decided to use high resolution everywhere and put our computational resources into a large unigrid:
$1024^3$ in this paper, and moving to $1536^3$ in recent work at $z \approx 6-10$ (Shull \etal\ 2012a).  

We use a modified version of the star-formation routine of Cen \& Ostriker (1992).  Star particles, 
representing the combined presence of a few million solar masses of stars, are formed when the 
following three conditions are met: the total density of a grid cell is above a certain threshold, a convergent 
flow exists (negative velocity divergence), and the cooling time is less than the dynamical time.
Star particles return feedback to the grid in the form of metal-enriched gas and thermal energy.  
We use a distributed-feedback method (Smith \etal\ 2011) in which material and energy are distributed 
over a 27-cell cube (three cells on a side) centered on the location of the star particle.  This is done to mitigate 
the over-cooling that occurs when feedback is deposited into a single grid cell, which raises the temperature,
gas density, and cooling rate to unphysically high values.  Over-cooling also causes the winds that should be 
transporting metals into the IGM to fizzle out and remain confined to their sites of origin.  Smith \etal\ (2011) 
showed that this model is able simultaneously to provide good matches to the global star formation history and 
the observed number density per unit redshift of \OVI\  absorbers (Danforth \& Shull 2008).  
If all three above conditions are satisfied, a ``star particle," representing a large collection of stars, is formed 
within the grid cell with a total mass, $m_{*} = f_{*}  m_{\rm cell} (\Delta t/t_{\rm dyn})$.  
Here, $f_{*} \approx 0.1$ is the star-formation efficiency, $m_{\rm cell}$ is the baryon mass in the cell, $t_{\rm dyn}$ 
is the dynamical time, and $\Delta t$ is the hydrodynamical timestep.  Subsequently, this much mass is also
removed from the grid cell as the star particle is formed, ensuring mass conservation.  Although the star particle 
is formed instantaneously within the simulation, feedback is assumed to occur over a longer time scale, which 
more accurately reflects the gradual process of star formation.   Stellar feedback is represented by the injection of
thermal energy and the return of gas and metals to the grid, in amounts proportional to $\Delta m_{\rm sf}$.  
We assume that a fraction (25\%) of the stellar mass is returned to the grid as gas.  The thermal energy and metals
returned to the grid are $e =  \epsilon (\Delta m_{\rm sf}) c^{2}$ and $m_{\rm metals} = (\Delta m_{\rm sf}) y$, where
$\epsilon \approx 10^{-5}$ is the ratio of rest-mass energy to thermal energy and $y \approx 0.025$ is the metal yield.  
The rate of star formation and rate at which thermal energy and metals are injected into the grid by the star particle 
peak after one dynamical time, then decays exponentially.   

In the current project, we seek to understand the thermal and ionization state of the IGM, including the  distribution 
and covariance of metallicity, temperature, and \OVI\ ionization fraction.   In our simulations, the physical 
properties of \OVI\ absorbers and the degree to which they trace the WHIM are in good agreement with 
other recent simulations (Tepper-Garcia \etal\  2011;   Cen \& Chisari 2011).  However, they differ significantly 
from those of Oppenheimer \& Dav\'e (2009, 2011), who found that \OVI\ and \NeVIII\  originate almost exclusively 
in warm ($T \approx 10^4$\,K) photoionized gas.  Tepper-Garcia \etal\ (2011) suggested that much of this 
difference arises because Oppenheimer \& Dav\'e (2009) neglected the effect of photoionization on metal-line cooling.
Oppenheimer \etal\ (2012) investigated this claim by running additional simulations in which photoionization 
reduced the metal cooling.  This reduction resulted in a small, but insufficient number of \OVI\ absorbers 
associated with WHIM gas.  Instead, they point to the fact that they do not include the mixing of metals from their 
galactic-wind particles, allowing feedback to take the form of cold, heavily enriched clouds.  

In examining the various interpretations of the high ions (\OVI\ and \NeVIII), we have identified several key differences 
between the two codes:  the Oppenheimer-Dav\'e smooth particle hydrodynamic (SPH) code and our grid-based
approach with \texttt{Enzo}.  The disagreement between the IGM  temperatures and ionization mechanisms appear to 
arise from four effects:  
(1)  Different methods of injecting energy and metals (mixed or unmixed);
(2)  Over-cooling of unmixed, metal-enriched gas at high density and high metallicity; 
(3)  Differences in shock capturing (and shock-heating) between SPH and grid codes;
(4)  Differences in photoionization rates through the assumed radiation fields at $h \nu = 100-250$ eV.
All four possibilities merit careful comparative studies, although we believe the resolution of this \OVI\ controversy may 
ultimately hinge on understanding the nature of  galactic winds and their ability to mix heavy elements into IGM.

\section{CENSUS OF BARYONS IN DIFFERENT THERMAL PHASES}


\begin{figure*}
   \includegraphics[width=0.9\textwidth]{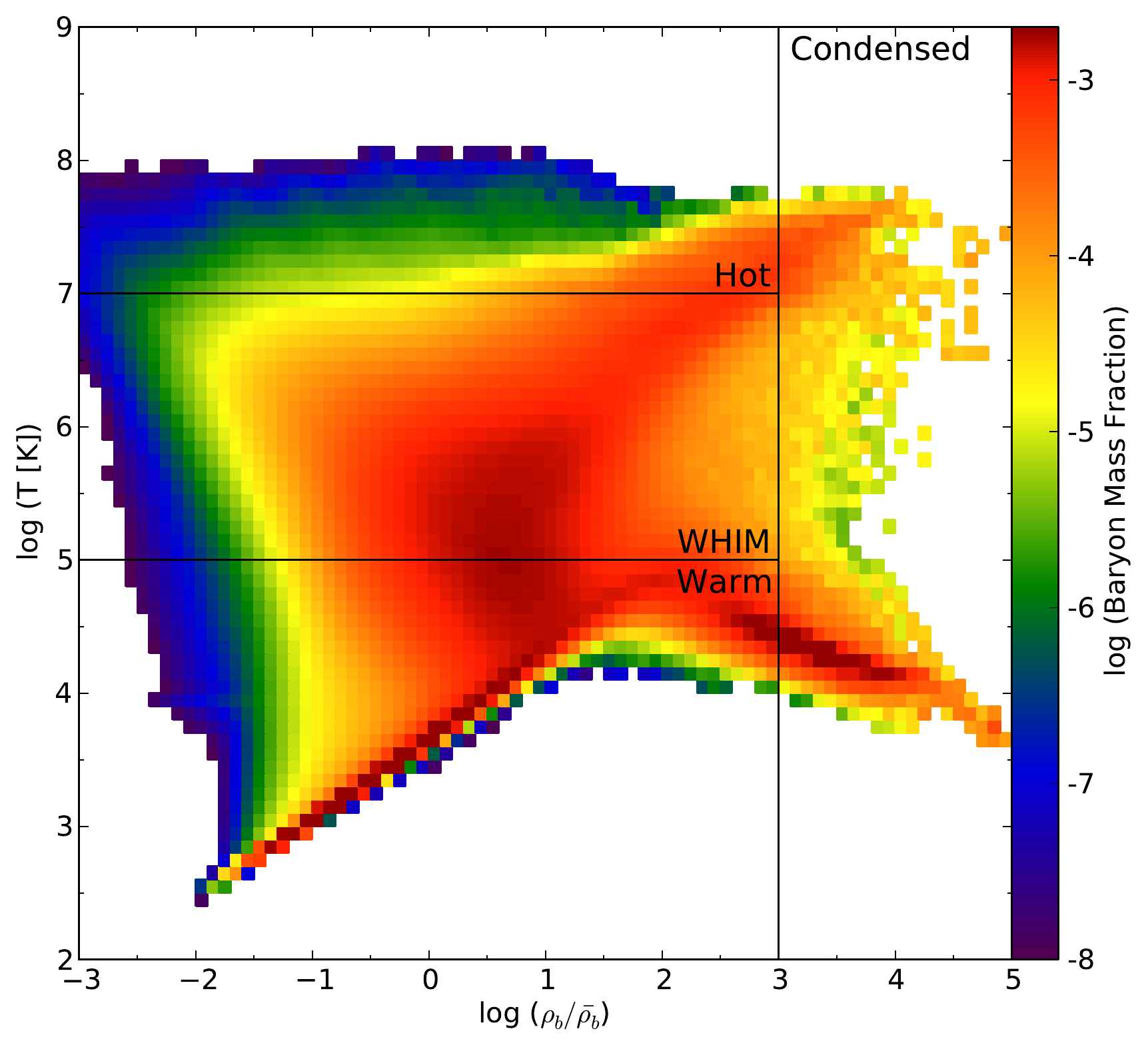}
   \caption{ Distribution of IGM in temperature $T$ and baryon overdensity 
   $\Delta_b = \rho_b / \overline{\rho}_b$, color-coded by baryon mass fraction.  
   This distribution shows the same thermal phases seen in simulations by other groups and
   commonly labeled as warm (diffuse photoionized gas), WHIM, and condensed.    
      }
  \end{figure*} 


The ``Lyman-$\alpha$ forest" of absorption lines appears to contain $\sim$30\% of the low-$z$ baryons
(Penton \etal\ 2000, 2004; Lehner \etal\  2007).  Another 30--40\% is predicted by simulations to reside in 
shock-heated gas at $10^5$~K to $10^7$~K (WHIM).  Owing to its low density, the WHIM is difficult to detect 
in emission (Soltan 2006).  More promising are absorption-line studies, using the high ionization states of abundant 
heavy elements with resonance lines in the far-ultraviolet (\CIV, \NV, \OVI),  extreme ultraviolet (\OIV, \OV, \NeVIII), 
and soft X-ray (\OVII, \OVIII, \NVI, \NeIX).   The low-$z$ WHIM has most effectively been surveyed in the \OVI\  lines 
at 1031.926~\AA\  and 1037.617~\AA\ (Danforth \& Shull 2005, 2008; Tripp \etal\ 2008; Thom \& Chen 2008), 
which probe the temperature range $10^{5.3-5.7}$~K in collisionally ionized gas.   Danforth \& Shull (2008) measured 
the column densities of 83 \OVI\  absorbers and estimated that $8.6\pm0.8$\% of the baryons reside in this phase, 
assuming constant correction factors for the metallicity ($Z  \approx 0.1\,Z_{\odot}$) and \OVI\ ionization fraction 
($f_{\rm OVI} = 0.2$).   A few detections of \NeVIII\ have also been reported (Savage \etal\ 2005;  Narayanan \etal\ 
2009, 2011; Meiring \etal\ 2012) probing somewhat hotter gas.    Weak X-ray absorption lines are difficult to detect 
with the current throughput and spectral resolution of spectrographs on {\it Chandra} and {\it XMM/Newton}.  Possible 
X-ray detections of hotter gas at $(1-3)\times10^6$~K have been claimed, using absorption lines of helium-like \OVII\ 
$\lambda21.602$ (Nicastro \etal\ 2005a,b, 2008;  Buote \etal\ 2009; Fang \etal\ 2010;  Zappacosta \etal\ 2010) and 
hydrogenic \OVIII\ $\lambda18.969$ (Fang \etal\ 2002, 2007).  Most of these {\it Chandra} detections remain 
controversial and unconfirmed by the {\it XMM-Newton} satellite (Kaastra \etal\ 2006; Williams \etal\ 
2006; Rasmussen \etal\ 2007).  Recent analyses of spectroscopic data on Mrk~421 fails to detect any WHIM gas at 
the claimed redshifts ($z = 0.01$ and 0.027), either in broad \Lya\ absorption (Danforth \etal\ 2011) from high-S/N 
data from the Cosmic Origins Spectrograph (COS) on the {\it Hubble Space Telescope} (\HST) or in \OVII\  
(Yao \etal\ 2012) in {\it Chandra} data.

The post-processed results of our simulations relevant to \OVI\  are illustrated in Figures 1--7.  Figure~1 shows 
the temperature-density phase diagram of the multiphase IGM, color-coded by baryon mass fraction.  The commonly 
found features in ($T, \Delta_b$) plots include the diffuse \Lya\ absorbers ($T = 10^{3.0-4.5}$~K and 
$\Delta_b = 10^{-2}$ to $10^{+1.5}$), a condensed phase ($\Delta_b > 10^3$), and a shocked-heated plume of 
WHIM  ($T = 10^{5-7}$~K and $\Delta_b = 10^{0}$ to $10^{2.5}$).   The gas traced by \OVI\  resides at 
$5 < \log T < 6$,  marked in Figure 2 for the cumulative distribution of mass and metals vs.\ temperature in our 
simulations.  The two curves show results from simulations with distributed feedback (energy injected into 27 cells) 
and local feedback (single cell).   The similarity of the mass distribution curves is evidence of model robustness.  
Figure~3 shows the cumulative mass distribution of \OVI\ versus column density, indicating that a significant  fraction 
of \OVI\ is traced by weak absorbers with $\log N_{\rm OVI} \leq 13.5$ (equivalent width $W_{\lambda} < 40$~m\AA\
for stronger 1032~\AA\ line).

Our simulations show a large range of \OVI\ properties, including metallicity,  temperature, and ionization fraction.  
Figures 4 and 5 show the range and covariance of the two correction factors:   metallicity in solar units $(Z / Z_{\odot})$ 
and \OVI\ ionization fraction $f_{\rm OVI}$.  The ionization fraction includes the effects of both collisional ionization and 
photoionization, as discussed by Smith \etal\ (2011).  Figure~4 color-codes these distributions in \NOVI, while Figure~5 
illustrates their distribution in temperature and baryon overdensity.  
The wide range of physical conditions in which \OVI\ exists demonstrates the multiphase character of this ionized gas.  
The mechanisms that produce \OVI\ (113.87~eV is needed to ionize \OV) include collisional ionization at $\log T >  5$ 
and photoionization by the hard (EUV) metagalactic background.  The dashed line in Figure~5 shows the locus of points 
($\log T$ and $\log \Delta_b$) at which collisional ionization equals photoionization.  This calculation is based on \OV\ 
collisional ionization rates from Shull \& Van Steenberg (1982).  The \OV\ photoionization rate was derived, assuming a
 cross section $\sigma (E) = (0.36 \times 10^{-18}~{\rm cm}^2)(E/113.87~{\rm eV})^{-2.1}$ and the EUV radiation field at 
 8--10 ryd from Figure 13 of Haardt \& Madau (2012).  

Figure 6 shows the distribution of the product, $(Z /Z_{\odot}) \, f_{\rm OVI}$, throughout the simulation.  Color-coded by
\OVI\ mass fraction, this plot exhibits bimodality of this product with temperature and baryon overdensity.  In the deepest-red 
portions, representing high \OVI\ mass fraction, one finds regions with large values, 
$(Z / Z_{\odot}) \, f_{\rm OVI} \approx 10^{-1.3}$, in high-density filaments with $\Delta_b \approx 100$, and other 
regions with lower values, $(Z /Z_{\odot}) \, f_{\rm OVI} \approx 10^{-2.5}$ at  $\Delta_b \approx 10$.  In converting 
the column densities of \OVI\ absorbers to baryons, one therefore should not adopt single corrections for metallicity and 
ionization fraction.   However, as we discuss in Section 3.1, one can apply statistical corrections that correlate the product 
$(Z /Z_{\odot}) \, f_{\rm OVI}$ with \OVI\ column density (Figure 7).  Integrating over the \OVI\ column-density distribution, 
we find that this product is a factor of two lower than the previously assumed value of 0.02.   This doubles the fraction of 
WHIM baryons traced by \OVI\ from 8--9\% to $17\pm4$\%.


\begin{figure}
  \includegraphics[width=0.5\textwidth]{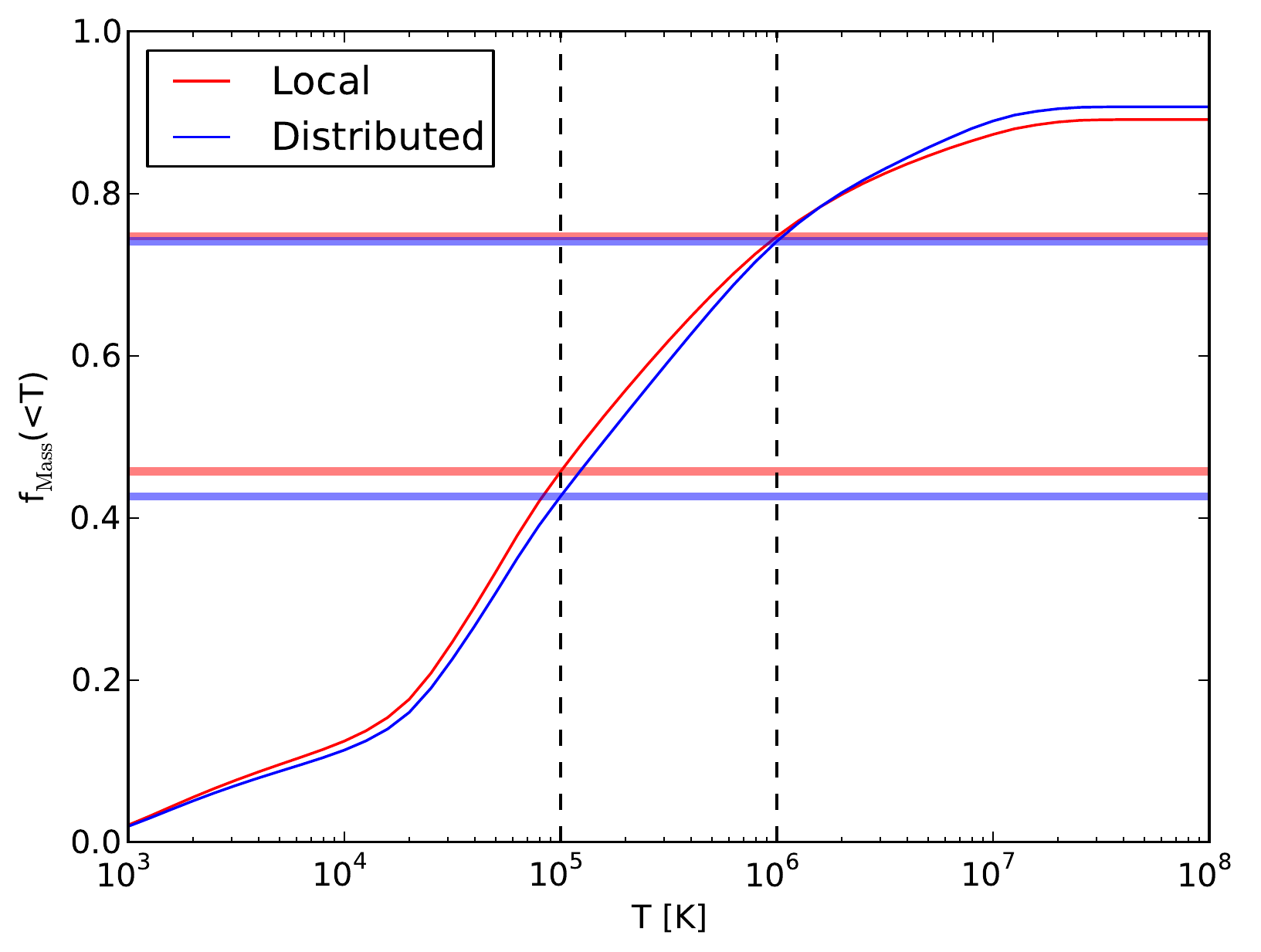} 
  \includegraphics[width=0.5\textwidth]{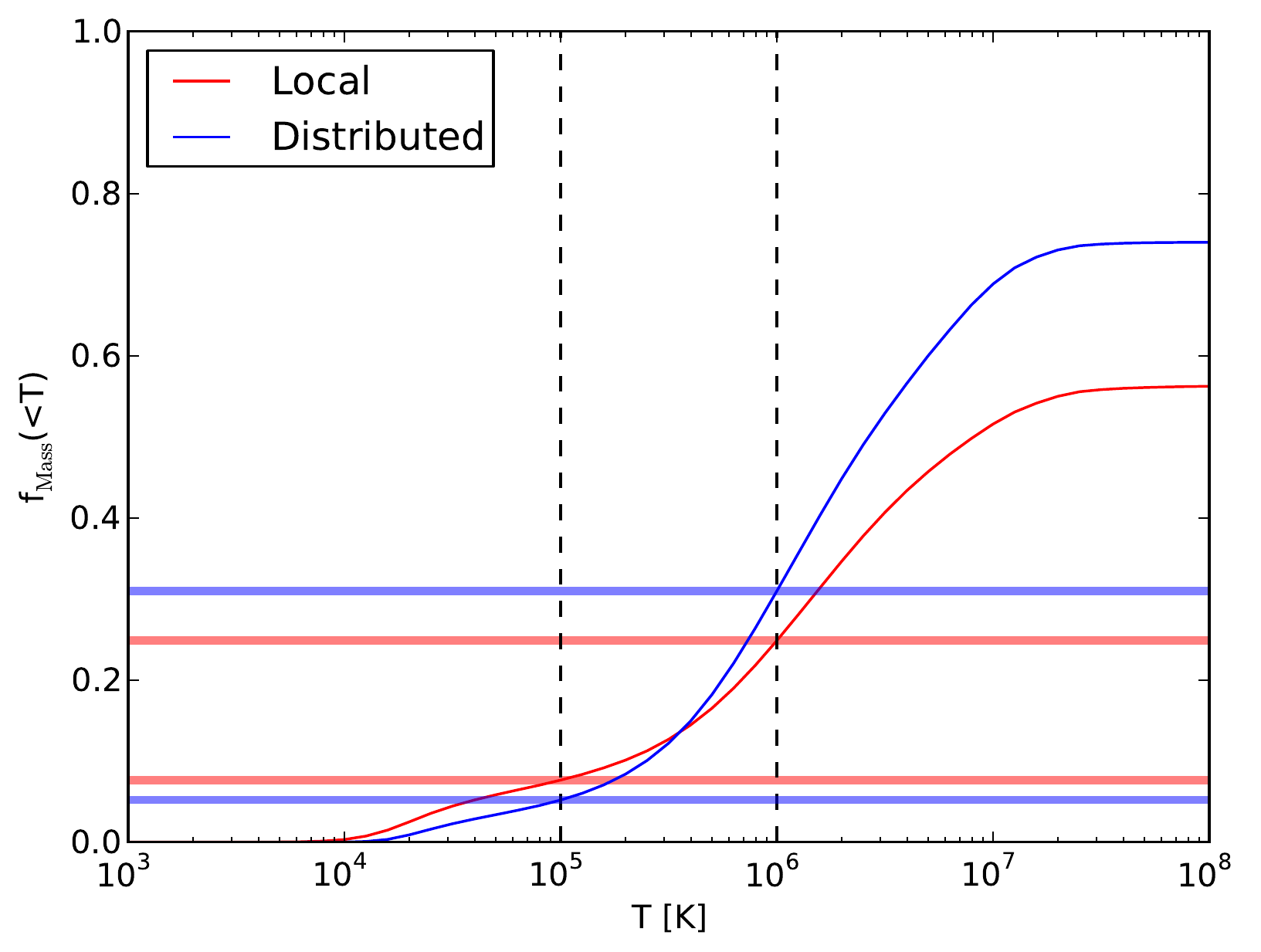} 
  \caption{Cumulative distributions of IGM mass (top panel) and metals  (bottom panel) 
       vs.\ temperature $T$ from our simulations, run with both ``local and distributed" feedback 
       (Smith \etal\ 2011).   We only show IGM baryons with overdensities $\Delta_b < 1000$; the
       remaining mass and metals are in galaxies and collapsed phase.  Vertical dashed lines mark the 
      temperature range ($10^5$~K to $10^6$~K) probed by \OVI\ and \NeVIII, and horizontal lines
      show intercepts for the two distributions.        }
  \end{figure} 



\begin{figure} 
  \includegraphics[width=0.5\textwidth]{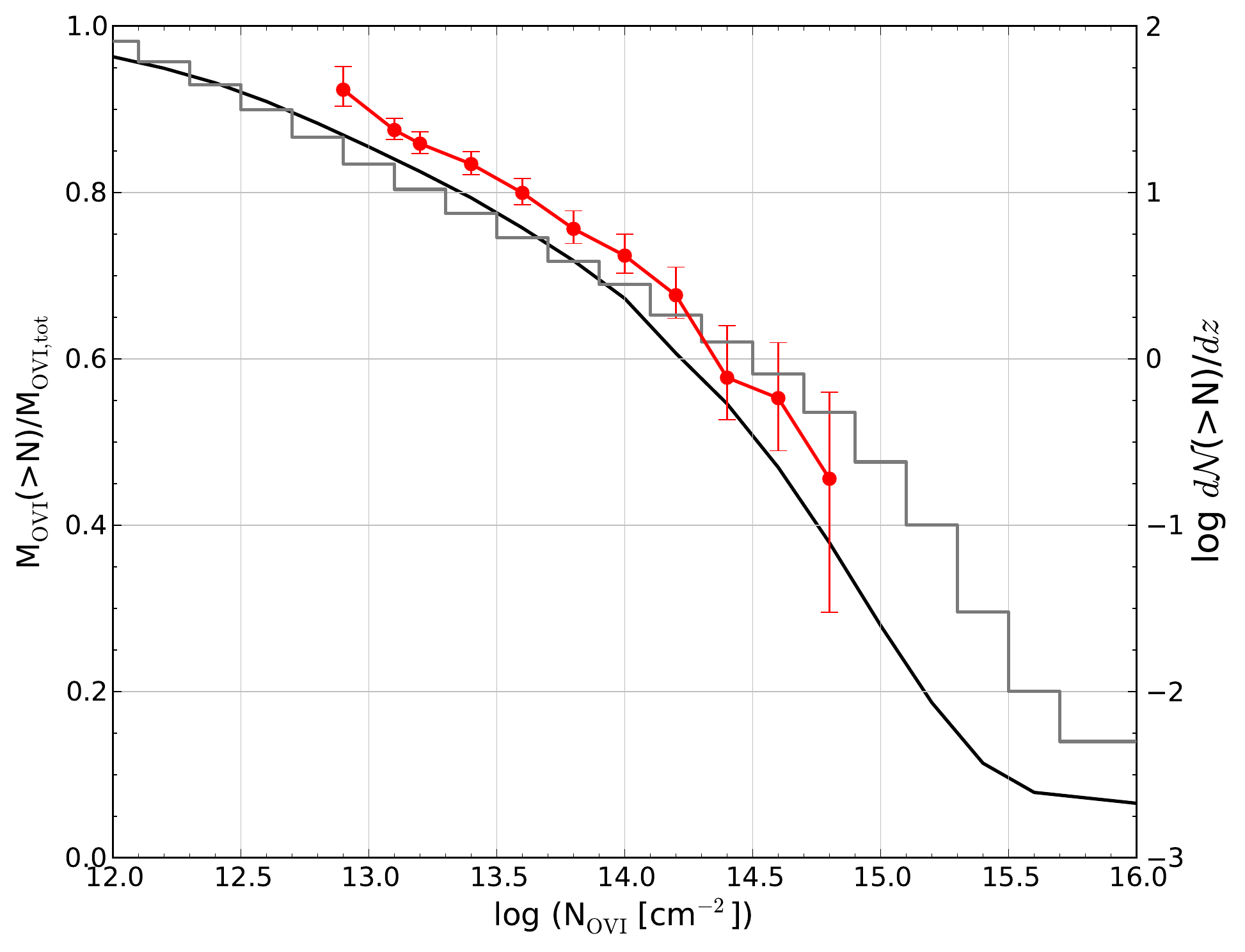}
   \caption{Cumulative distribution of \OVI\ by mass from our WHIM simulation,
   for column densities $N_{\rm OVI} = 10^{12}$ to $10^{16}$ \cd.  
   Solid black line shows cumulative mass fraction, labeled on left vertical axis.
   Gray histogram shows cumulative number of \OVI\ absorbers, labeled on right vertical axis.   
   Red points show observed cumulative column-density 
   distribution (Danforth \& Shull 2008) labeled on right vertical scale.
   A significant fraction of \OVI-traced WHIM resides in weak absorbers, although
   the distribution flattens at $\log N_{\rm OVI} < 13.5$.          }
  \end{figure} 



\begin{figure}
 \includegraphics[width=0.5\textwidth]{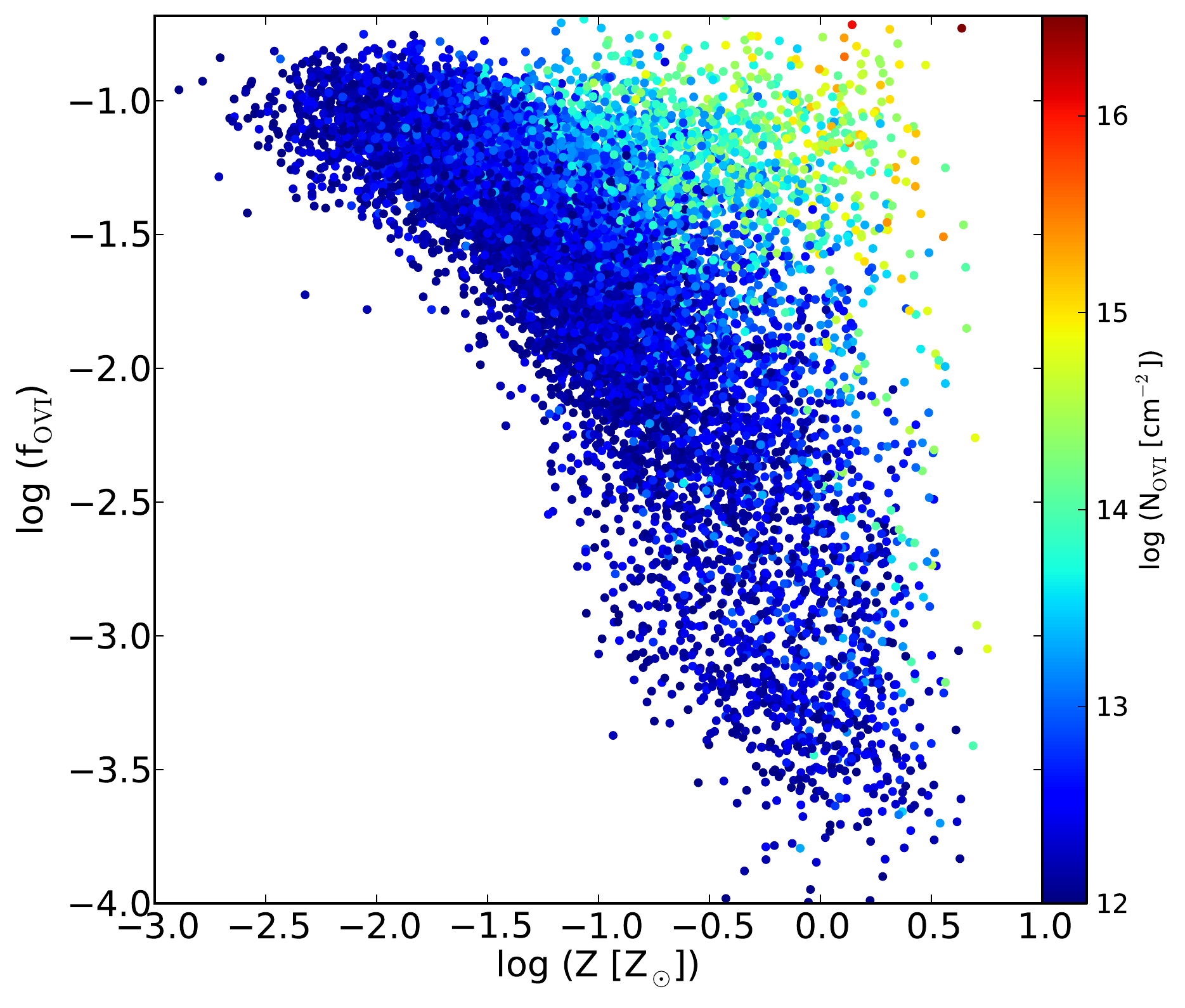}
   \caption{Distribution of IGM metallicity $(Z/Z_{\odot})$ and  \OVI\ ionization fraction $(f_{\rm OVI})$ 
   color-coded by \OVI\ column density.  Note the wide range and covariance of individual factors, 
   whose product, $f_{\rm OVI} \,(Z/Z_{\odot})$, correlates with \OVI\ column density (color bar along right).   
      }
  \end{figure} 



\begin{figure*}
 \includegraphics[width=0.95\textwidth]{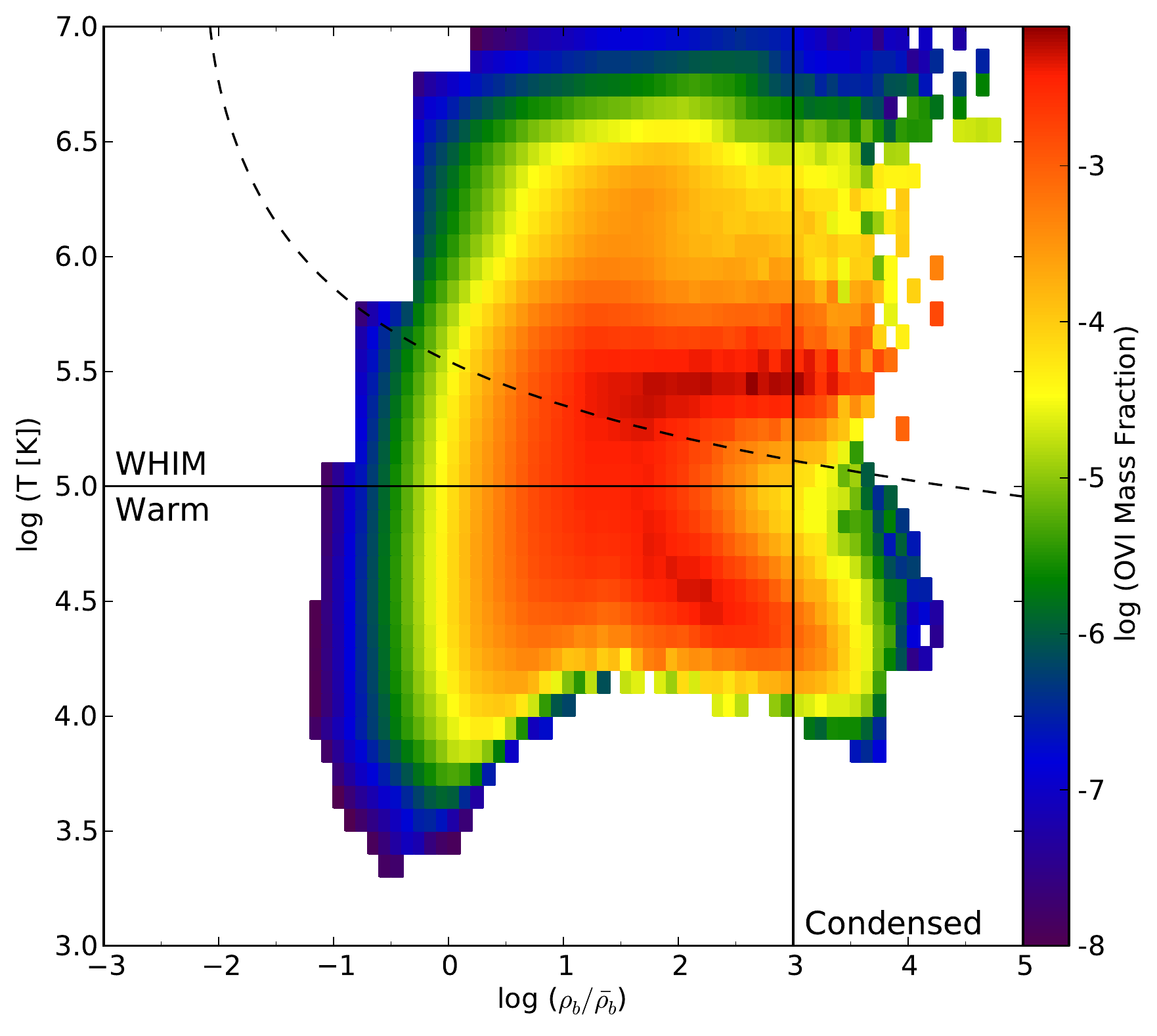}
   \caption{Distribution of IGM temperature versus baryon overdensity, $\Delta_b = \rho_b / \overline{\rho}_b$, 
   color-coded by \OVI\ mass fraction.   In this phase space, we identify the WHIM ($T \geq 10^5$~K), warm, diffuse 
   photoionized gas ($T < 10^5$~K and $\Delta_b < 1000$), and condensed gas  ($\Delta_b > 1000$).  Dashed line 
   shows locus at which collisional ionization equals photoionization (collisional ionization dominates
   above the dashed curve). 
        }
  \end{figure*} 



\begin{figure}
 \includegraphics[width=0.5\textwidth]{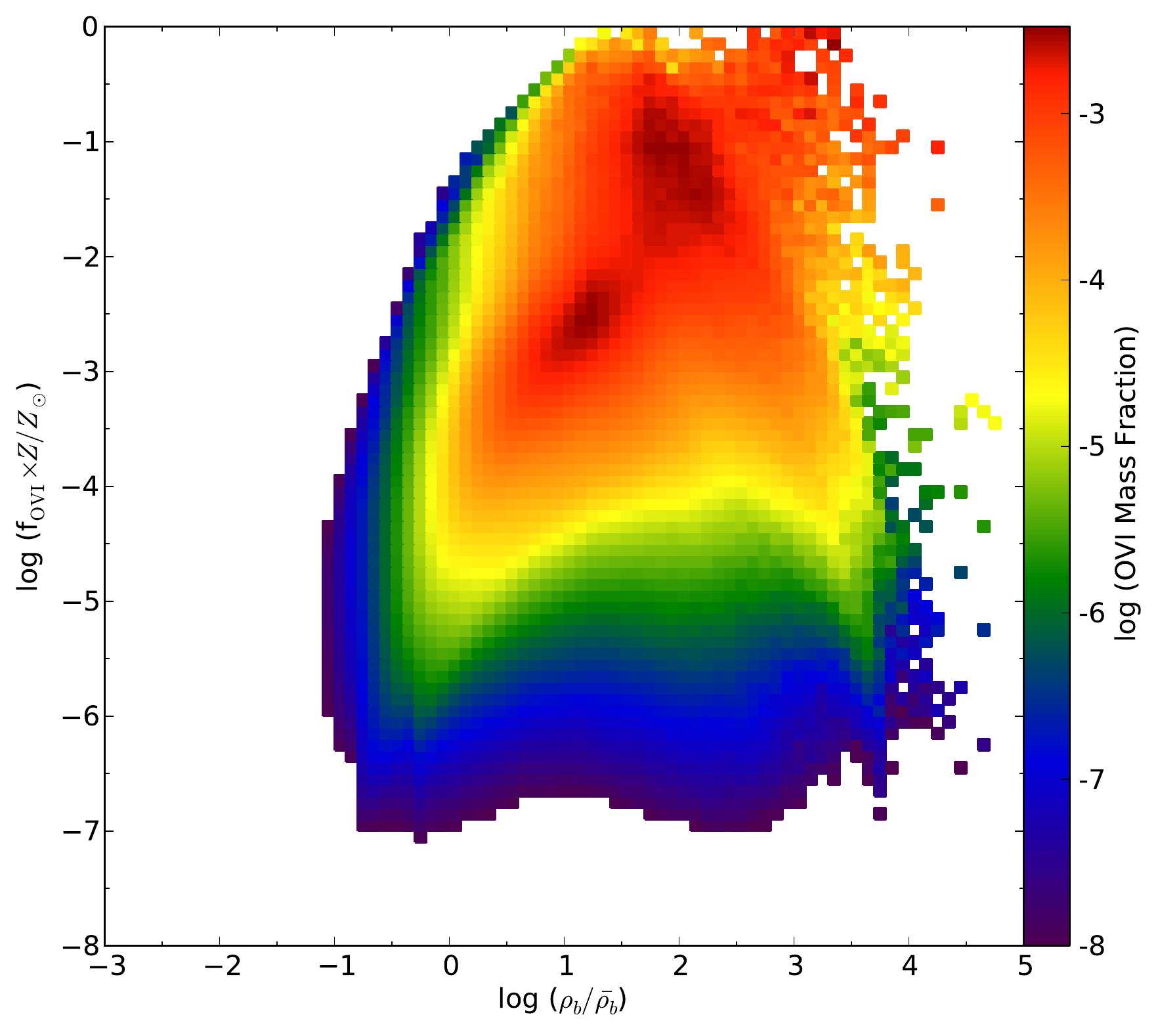}
   \caption{Distribution of the metallicity-ionization product, $(Z /Z_{\odot}) \, f_{\rm OVI}$,
   versus baryon overdensity, $\Delta_b = \rho_b / \overline{\rho}_b$, color-coded by \OVI\ mass
   fraction.   The broad distribution of the product, from 0.001 to 0.1 has local enhancements 
   (deep red) in low-metallicity regions (overdensities $\Delta_b \approx 10$) and high-metallicity 
   regions ($\Delta_b \approx 100$).  The column density weighted mean of this product is 0.01 (Figure 7). 
           }
  \end{figure} 



\begin{figure*}
\includegraphics[width=0.8\textwidth]{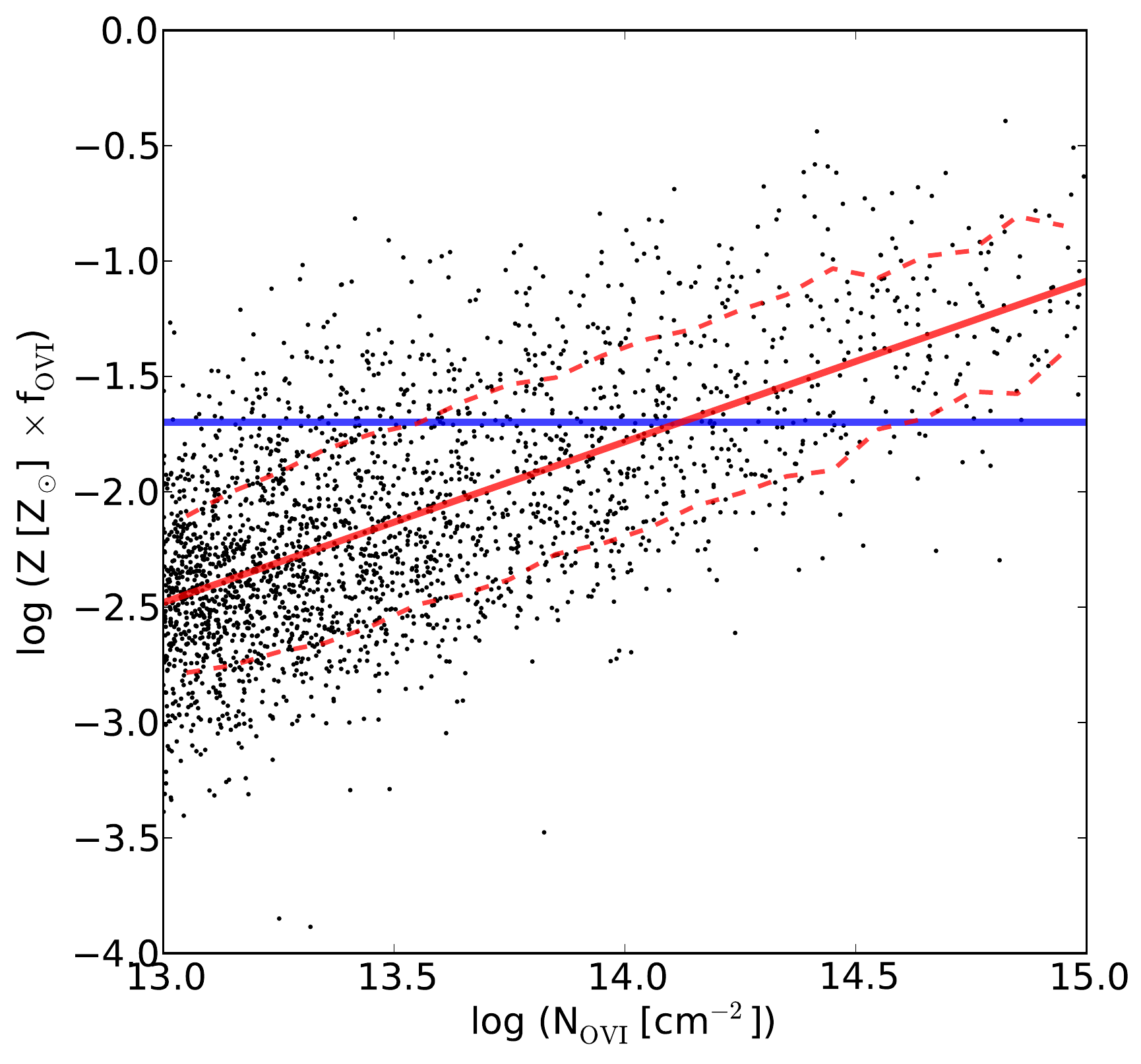}
   \caption{Values of the product, $(Z /Z_{\odot}) \, f_{\rm OVI}$, for the WHIM simulation with
    distributed feedback overlaid with a power-law fit (solid red line) to the ensemble, 
    $(0.015)[N_{\rm OVI} / 10^{14}~{\rm cm}^{-2}]^{0.70}$.  Results from simulations with local feedback
    are similar.  
    Integrating over the observed distribution in \NOVI, we find a weighted value of 0.01.  Horizontal (blue) 
    line shows previously assumed value of 0.02 for this product ($Z/Z_{\odot} = 0.1$ and $f_{\rm OVI} = 0.2$). 
     Typical variance ($1\sigma$) about this fit is 0.3--0.4 dex, shown by red dashed lines.  
      }
  \end{figure*} 


\subsection{Warm-Hot IGM Probed in \OVI\ Absorption}  

As noted, the metallicity and \OVI\ ionization fraction vary throughout the grid in the simulated IGM.  
Figures 6 and 7 show that the product $(Z/Z_{\odot}) f_{\rm OVI}$  correlates with \NOVI, 
with weaker absorbers having systematically lower values.   This is primarily an effect of spatial
variations in the metallicity, but also variations in the \OVI\ ionization fraction produced
by spatial fluctuations in WHIM temperature and contributions from  photoionizing radiation
on low-density IGM.   The baryon content  in \OVI-traced WHIM is given by an integral over \OVI\ 
column density ($N$),  now with the corrections for metallicity and ionization fraction placed
inside the integral:
\begin{equation}
  \Omega_b^{{\rm (OVI)}} = \left[  \frac {\mu_b H_0} {c \rho_{\rm cr} {\rm (O/H)}_{\odot}}   \right]
                           \int_{N_{\rm min}}^{N_{\rm max}} \frac {d {\cal{N}} (N)} {dz} \, 
                            \frac  {N \; dN } { Z_{\rm O} (N) \, f_{\rm OVI}(N) }    \; . 
\end{equation} 
Here, $\mu_b = 1.33 m_H$ is the mean baryon mass per hydrogen nucleus, accounting for helium.  
Earlier surveys that used \OVI\ as a baryon tracer assumed constant values of  the ionization fraction,  
$f_{\rm OVI} = 0.2$ (its maximum value in collisional ionization equilibrium at $\log T_{\rm max} = 5.45$) 
and metallicity, $Z /Z_{\odot} = 0.1$, relative to the solar oxygen abundance, 
(O/H)$_{\odot} = 4.90 \times 10^{-4}$ (Asplund \etal\ 2009).  

From our simulations (Figure 7), the product of metallicity and \OVI\  ionization fraction has a 
statistical power-law dependence on \OVI\ column density, scaling as  $N^{\gamma}$,
\begin{equation}
   (Z/Z_{\odot}) \, f_{\rm OVI} = (0.015)(N_{\rm OVI} /10^{14}\,{\rm cm}^{-2})^{0.70}  \;  .
\end{equation}
Integrating over the distribution $d{\cal{N}} / dz \propto N^{-\beta}$, we find
\begin{equation}
  \Omega_b^{{\rm (OVI)}} \propto  \int_{N_{\rm min}}^{N_{\rm max}}  \, N^{1 - \gamma - \beta} \; dN 
     \propto [ N_{\rm min}^{2-\gamma-\beta} - N_{\rm max}^{2-\gamma-\beta}  ] \; .  
\end{equation} 
For $\gamma \approx 0.70$ and the observed $\beta \approx 2.0$ (Danforth \& Shull 2008),  the integral 
increases at the low end as $N_{\rm min}^{-0.7}$.  The column-density weighted mean of  the 
product $f_{\rm OVI} \, (Z / Z_{\odot}) = 0.01$, a factor of two smaller than previously assumed.  
For our standard integration range, $13.0 \leq \log N_{\rm OVI} \leq 15.0$, this correction doubles 
the number of baryons in the \OVI-traced  WHIM, compared to previous assumptions.  As noted earlier, 
we found similar results in all our simulations. 

Previous \OVI\ surveys and baryon estimates were quantified by an absorption-line frequency, $d{\cal N}/dz$, 
per unit redshift and an \OVI-traced baryon fraction, $\Omega_b^{(\rm OVI)} / \Omega_b$.  Using 40 \OVI\ absorbers 
seen with \FUSE, Danforth \& Shull (2005)  found $d{\cal N}/dz \approx 17\pm3$ for column densities 
$13.0 \leq N_{\rm OVI} \leq 14.5$.  The $10^{13}$~\cd\ lower limit corresponds to 12.5~m\AA\ equivalent width 
in \OVI\ $\lambda1032$.  The 2005 census gave an \OVI-traced baryon fraction of at least $4.8\pm0.9$\% (statistical 
error only) for a correction factor $f_{\rm OVI} \, (Z / Z_{\odot}) = 0.02$.  Danforth \& Shull (2008) used \HST/STIS 
data on 83 \OVI\ absorbers to find $d{\cal N}/dz \approx 15.0^{+2.7}_{-2.0}$ integrated to 30 m\AA, with
$d{\cal N}/dz \approx 40^{+14}_{-8}$ integrated to 10 m\AA.  Their derived WHIM baryon fractions 
were $7.3\pm0.8$\% and $8.6\pm0.8$\%, integrated to 30~m\AA\ and 10 m\AA, respectively.  
Tripp \etal\ (2008) found a similar line frequency, $d{\cal N}/dz \approx 15.6^{+2.0}_{-2.4}$ 
for 51 intervening \OVI\ absorption systems integrated to 30 m\AA, while Thom \& Chen (2008) found 
$d{\cal N}/dz \approx 10.4\pm2.2$ for 27 \OVI\ absorbers down to 30 m\AA.   The latter survey was more 
conservative, requiring detection of both the stronger \OVI\ line at 1032~\AA\ and the weaker line at 1038~\AA.
They also had fewer sight lines  (16) and a much smaller number of  \OVI\ absorbers (27) in their survey.  
Their line frequency is smaller than those in other \OVI\  surveys, perhaps because of small-number statistics
or owing to the two-line requirement.  

As part of a Hubble Archive Legacy project (PI:  Shull, HST-AR-11773.01-A), the Colorado group has reanalyzed
 \HST/STIS data on IGM absorption lines (Tilton \etal\  2012), finding 746 \HI\ absorbers and 111 \OVI\ absorbers.  
Through a critical evaluation of data in the literature and comparison to high-S/N COS spectra when available, we 
corrected line identification errors in Danforth \& Shull (2008) and other surveys.   The \OVI\ line frequency from 
Tilton \etal\ (2012) is $d{\cal N}/dz \approx 22.2^{+3.2}_{-2.4}$ integrated to 30 m\AA, with a slope 
$\beta = 2.075 \pm 0.119$.  This new survey provides more reliable baryon fractions for the \OVI-traced WHIM of 
$7.2\pm0.8$\% (integrated down to 30 m\AA\ equivalent width) and $8.6\pm0.7$\% (to 10 m\AA), using the old 
correction factor, $f_{\rm OVI} \, (Z / Z_{\odot}) = 0.02$.  Taken as a whole, these \OVI\ surveys suggest a 
WHIM baryon fraction of 8--9\%, assuming the old values for metallicity and ionization fraction.
 With our new correction factors, $f_{\rm OVI} \, (Z / Z_{\odot}) = 0.01$, the baryon fractions double.  We therefore 
 adopt an \OVI-traced baryon fraction of $17\pm4$\%, where we have increased the error to account for systematic 
 uncertainties.  The value of $\Omega_b^{(\rm OVI)}$ is dominated by weak absorbers with $N_{\rm OVI} < 10^{13.5}$~\cd, 
where the absorption-line statistics become uncertain.  An accurate \OVI\ census will require measuring even weaker
\OVI\  $\lambda1032$ absorption lines, with column densities below $10^{13}$~\cd, corresponding to 
equivalent widths $W_{\lambda} = (12.5~{\rm m\AA})(N_{\rm OVI}/10^{13}$~\cd).   Deep spectroscopic surveys 
in \OVI\ can also ascertain where the distribution of \OVI\ absorbers flattens (Figure 3).

\subsection{Diffuse Photoionized \Lya\ Filaments}  

In this subsection, we provide details on the assumptions, described briefly in Penton \etal\ (2000), for the 
photoionization corrections to the \HI\ (\Lya) absorbers.   In photoionization equilibrium, the density of neutral 
hydrogen in low-density gas  depends on the ionizing background, gas density, and electron temperature, all 
of which evolve with redshift and have spatial fluctuations throughout the IGM: 
\begin{equation}
    n_{\rm HI} = \frac {n_e \, n_H \, \alpha_H^{(A)}(T) } { \Gamma_H }  \;  .
\end{equation}
Here, $n_e = (1 + 2y) = 1.165 n_H$ is the electron density for fully ionized gas, $n_H$ is the total density of 
hydrogen nuclei, and $y = n_{\rm He} / n_{\rm H} = [(Y/4)/(1-Y)] \approx 0.0823$ is the helium-to-hydrogen ratio 
by number, assuming $Y = 0.2477$ helium abundance by mass.   For the low values of $n_H$ and N$_{\rm HI}$
in IGM absorbers, we adopt the hydrogen case-A radiative recombination rate coefficient, 
$\alpha_H^{(A)}(T) \approx (2.51 \times 10^{-13}~{\rm cm}^3 \; {\rm s}^{-1}) T_{4.3}^{-0.726}$,
scaled to an electron temperature $T = (10^{4.3}~{\rm K}) T_{4.3}$ characteristic of low-metallicity IGM 
(Donahue \& Shull 1991).   The \HI\ photoionization rate depends on the metagalactic radiation field, with specific 
intensity $I_{\nu} = I_0 (\nu / \nu_0)^{-\alpha_s}$ referenced to the Lyman limit, $h\nu_0 = 13.6$~eV.  Here,  
$\langle \alpha_s \rangle \approx 1.6-1.8$ is the mean QSO spectral index between 1.0--1.5 ryd (Telfer \etal\ 2002;
Shull \etal\ 2012b).  The frequency-integrated photoionization rate is given by the approximate formula,
$\Gamma_H   \approx    [4 \pi I_0  \sigma_0 /  h  (\alpha_s + 3)]  \approx (2.49 \times 10^{-14}~{\rm s}^{-1})  \,   
I_{-23}  (4.8 /  \alpha_s + 3)$,  where $I_0 = (10^{-23}$ erg~cm$^{-2}$~s$^{-1}$~Hz$^{-1}$~sr$^{-1}) I_{-23}$.
Hereafter, we combine the two parameters, $I_0$ and $\alpha_s$, into a single scaling parameter, 
$\Gamma_{-14} = (\Gamma_H / 10^{-14}~{\rm s}^{-1})$, for the hydrogen photoionization rate.
Previous calculations of the low-$z$ ionizing intensity (Shull \etal\ 1999) found 
$\Gamma_H \approx 3.2^{+2.0}_{-1.2} \times 10^{-14}~{\rm s}^{-1}$ and noted that low-$z$ observational 
constraints were consistent with values in this range.   More recent calculations (Haardt \& Madau 2012) 
are consistent with this value and other estimates at $z \approx 0$.   Figure 8 illustrates current estimates of the 
intensity of the metagalactic ionizing radiation field starting at 1 ryd (\HI\ photoionization edge), continuing to  
4 ryd (\HeII\ edge) and beyond, including ionization potentials needed to produce some of the higher metal 
ions, such as \CIV\ (47.87 eV = 3.52 ryd), \OVI\ (113.87 eV = 8.37 ryd), \OVII\ (138.08 eV = 10.15 ryd), 
\NeVIII\ (207.2 eV = 15.24 ryd), and \OVIII\ (739.11 eV = 54.35 ryd).


\begin{figure}
 \includegraphics[width=0.5\textwidth]{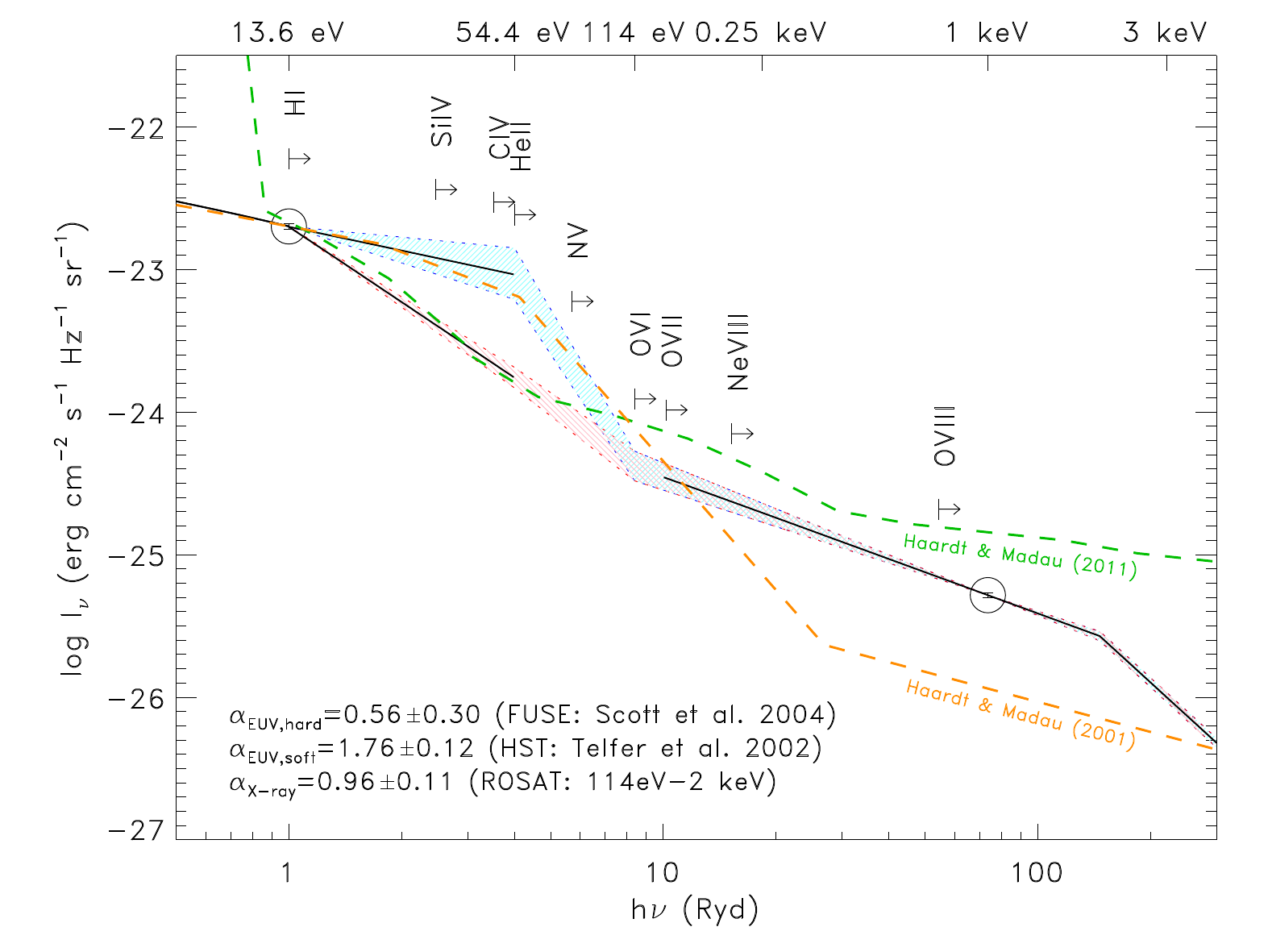}
    \caption{Compilation of the mean AGN spectral energy distribution (SED), pinned at 1 ryd, with
   $I_0 = 2 \times 10^{-23}$ erg~cm$^{-2}$~s$^{-1}$ Hz$^{-1}$ sr$^{-1}$ and two different flux
   distributions, $F_{\nu} \propto \nu^{-\alpha_s}$.  Previous composite spectra from small numbers 
   of AGN are shown in red (HST) and light blue (FUSE), with slopes $\alpha_s = 1.76 \pm 0.12$ 
   (Telfer \etal\ 2002) and $\alpha_s = 0.56^{+0.38}_{-0.28}$ (Scott \etal\ 2004).   Soft X-ray data are 
   from ROSAT observations of AGN in the Lockman Hole (Hasinger 1994) normalized at $E = 1$ keV 
   (73.53 ryd) with spectral slope $F_E \propto E^{-0.96 \pm 0.11}$.  The ionization potentials of \HI\
   and  \HeII\ and ionization energies to produce key metal ions are shown as arrows.
   The metagalactic backgrounds are from Haardt \& Madau (2001, 2011). 
    }
  \end{figure} 


The \Lya\ absorbers are thought to arise as fluctuations in dark-matter confined clumps or filaments,
which we approximate as singular isothermal spheres with density profiles,
$n_H(r) = n_0 (r/r_0)^{-2}$ normalized to a fiducial radius $r_0$.  For a sight line passing through
an absorber at impact parameter $p$, the total baryon mass within radius $r = p$ is
$M_b(p) = 4 \pi \mu_b n_0  r_0^2  p$,  where $\mu_b = (1 + 4y) m_H \approx 1.33 m_H$ is the mean 
baryon mass per hydrogen.  The \HI\ column density can be derived, using the density-squared 
dependence of $n_{\rm HI}$ and integrating through the cloud along pathlength $\ell$ at impact 
parameter $p$, where $\ell^2 = r^2 - p^2$.  We substitute $p = r \cos \phi$ and $\ell = p \tan \phi$ for 
the angle $\phi$ between the directions of $p$ and $r$,
\begin{eqnarray}
   N_{\rm HI} (p) &=&   \left[  \frac  { \alpha_H^{(A)} \, (1 + 2y )}  { \Gamma_H } \right]  2 
        \int_{0}^{\infty}  n_0^2 \left( \frac {r} {r_0} \right) ^{-4}  \; dl   \nonumber \\
       &=&  \left[   \frac  { \pi n_0^2 \, r_0^4 \,  \alpha_H^{(A)} \, (1 + 2y )}  {2 \,  \Gamma_H \, p^3 } \right]   \; .
\end{eqnarray}         
Solving for the quantity $n_0 r_0^2$, we can express the absorber mass within $r = p$ as
\begin{eqnarray}
   M_b(p) &=&  4 \pi \mu_b \,  p^{5/2} \left[ \frac {2 \,  \Gamma_H \, N_{\rm HI}(p) } 
      { \pi (1 + 2y ) \alpha_H^{(A)} } \right]^{1/2}    \nonumber  \\
                 &=& (1.09 \times 10^9 \, M_{\odot}) \left[  N_{14}^{1/2} \, \Gamma_{-14}^{1/2} \, 
       p_{100}^{5/2} T_{4.3}^{0.363} \right]  \; .
\end{eqnarray}
In the above formula, we have scaled the impact parameter $p = (100~{\rm kpc}) p_{100}$
using a 100-kpc characteristic scale length of \Lya\ absorbers at fiducial column density
$N_{\rm HI} = (10^{14}\;{\rm cm}^{-2}) N_{14}$;  see Penton \etal\ (2000) and references therein.  

Next, we calculate the baryon density in the \Lya\ absorbers as a fraction of the cosmological closure 
density, $\rho_{\rm cr} = (3 H_0^2/8 \pi G)$ at $z = 0$.  Because our \Lya\ survey extends to $z = 0.4$, 
we include corrections for cosmological evolution in the space density of absorbers, 
$\phi(z)$, and the hydrogen photoionization rate, $\Gamma_H (z)$.  We begin with the standard 
expression for the number of absorbers per unit redshift,
\begin{equation} 
    \frac {d{\cal N}} {dz} = \left[ \frac {c}{(1+z) H(z)} \right]  \pi [p(z)]^2 \, \phi (z)  \; , 
\end{equation}
where $H(z) = H_0 [\Omega_m (1+z)^3 + \Omega_{\Lambda}]^{1/2}$ is the Hubble parameter
at redshift $z$ in a flat $\Lambda$CDM cosmology, and the absorber space density is 
$\phi(z) = \phi_0 (1+z)^3$.  The absorption-line frequency and impact parameter $p$ correspond 
to the \HI\ column density and absorber mass given in Equations (5) and (6).  The baryon mass 
density in \Lya\ absorbers at redshift $z$ is written as the product of absorber mass, $M_b$,
times the absorber space density, $\phi(z)$, integrated over the distribution in \HI\ column density.  
We define the closure parameter, $\Omega_b^{(\rm HI)} = \rho_b(0) / \rho_{\rm cr}(0)$
at $z = 0$, where $\rho_b(z) = \rho_b(0) (1+z)^3$.   Assembling all the terms, we find a closure 
parameter in  \Lya\ absorbers of column density \NHI,  
\begin{eqnarray}       
      \Omega_b^{(\rm HI)} &=& \frac { \phi_0 M_b } {\rho_{\rm cr}}     \nonumber  \\
      &=&   \frac {32 (2 \pi)^{1/2}} {3} \left[ \frac {\mu_b G}{c H(z) (1+z)^2} \right]
         \left[ \frac { \Gamma_H(z) p(z) N_{\rm HI} } { (1+2y) \alpha_H^{(A)} } \right] ^{1/2} \; . 
 \end{eqnarray}  
Recent calculations of the metagalactic ionizing background (Haardt \& Madau 2012) show that
the hydrogen photoionization rate rises rapidly from redshifts $z = 0$ to $z = 0.7$.  We have fitted the
rate to the convenient formula $\Gamma_H(z) = (2.28 \times 10^{-14}~{\rm s}^{-1})(1+z)^{4.4}$.  We have 
little data on the redshift evolution of the characteristic absorber scale length, $p$, other than theoretical 
expectations for  the gravitational instability of filaments in the cosmic web.  In the following calculation, 
we assume that  $p = (100~{\rm kpc}) p_{100}$ remains constant with redshift.  
We rewrite Equation (8) as an integral over column density,
 \begin{eqnarray}
   \Omega_b^{(\rm HI)} &=& (9.0 \times 10^{-5}) \frac { h_{70}^{-1} p_{100}^{1/2} \; T_{4.3}^{0.363} (1+z)^{0.2}}
           {  [\Omega_m (1+z)^3 + \Omega_{\Lambda}]^{1/2} }  \nonumber  \\
         & \times &  \int_{N_{\rm min}}^{N_{\rm max}} \frac { d {\cal N} (\log N_{\rm HI}) } {dz} 
             N_{14}^{1/2} \; d(\log \, N_{\rm HI})    \; . 
\end{eqnarray}        
The ionization rate, $\Gamma_{H}(z) \propto (1+z)^{4.4}$, enters Equation (8) as the square-root, 
almost exactly compensating for the $(1+z)^2$ factor in the denominator.  

From UV spectrographic surveys of intergalactic \Lya\ absorbers, we now have a reasonable understanding of 
the distribution of \HI\ column densities in the diffuse \Lya\ forest (Penton \etal\ 2000, 2004; Danforth \& Shull 2008;
 Lehner \etal\ 2007; Tilton \etal\ 2012).   Penton \etal\ (2004) applied photoionization corrections to a survey of 187 
 \Lya\ absorbers over redshift pathlength $\Delta z = 1.157$ and found a distribution in \HI\ column density 
 $f(N) \propto N^{-\beta}$ with $\beta = 1.65 \pm 0.07$ and a $29\pm4$\% contribution to the total baryon content.  
 Danforth \& Shull (2008) surveyed 650 \Lya\ absorbers with \HST/STIS over the range 
$12.5 \leq \log N_{\rm HI} \leq 16.5$ with total pathlength $\Delta z = 5.27$.  Their \HI\ distribution was similar, 
with $\beta = 1.73 \pm 0.04$ and a fractional contribution of  $28.7\pm3.7$\% (statistical errors only) to the 
baryon census.  Although the formulae for baryon fractions were stated correctly (Equations 7--10 in Danforth \& 
Shull 2008), the values listed in their Tables 12 and 13 were computed with incorrect cosmological corrections and 
overestimated values of $\Omega_b$ for \HI\ and \OVI.  Our new \HST\ archive survey (Tilton \etal\ 2012) properly 
includes these effects, as well as the redshift evolution of $\Gamma_H(z)$, to find $\beta = 1.68\pm 0.03$ for
746 \Lya\ systems with column densities between $12.5 <  \log N_{\rm HI} < 16.5$ over pathlength 
$\Delta z = 5.38$. Their new derivation of  $\Omega_b^{(\rm HI)}$ is consistent with 24-30\% of the baryons 
residing in the  \Lya\ forest and partial Lyman-limit systems.  For the current census, we adopt a baryon fraction of  
$\Omega_b^{\rm (HI)} = 28 \pm 11$\%.  The latter error bars include systematic effects, discussed in
greater detail in Appendix A.

\section{BARYON CENSUS AND FUTURE SURVEYS}

We now summarize the current status of the low-$z$ baryon census and discuss requirements
for future surveys.  Details of the individual baryon contributions are described in Appendix B,
giving previous estimates and uncertainties. 
Figure 9 shows a pie chart of the current observable distribution of low-redshift baryons in various 
forms, from collapsed structures to various phases of the IGM, CGM, and WHIM.   These slices show the
contributions, $\Omega_b^{(i)} / \Omega_b^{(\rm tot)}$, to the total baryon content from
components ($i$).    Measurements of \Lya, \OVI, and broad \Lya\ absorbers,
together with more careful corrections for metallicity and ionization fraction, can now account for
$\sim60$\% of the baryons in the IGM.  An additional 5\% may reside in circumgalactic gas, 7\% 
in galaxies, and 4\% clusters.  This still leaves a substantial fraction, $29\pm13$\%, unaccounted for.   
We have assigned realistic errors on each of the ``slices of the baryon pie", most of which involve systematic 
uncertainties in the parameters needed for the ionization corrections, metallicity, and geometric factors 
(cloud size).  It is possible that the baryon inventory could change as a result of better determinations 
of these parameters.   However, most numerical simulations including ours (Figure 2) suggest that a 
substantial reservoir ($\sim 15$\%) of hot baryons exists in the hotter WHIM ($T > 10^6$~K).


\begin{figure}
 \includegraphics[width=0.5\textwidth]{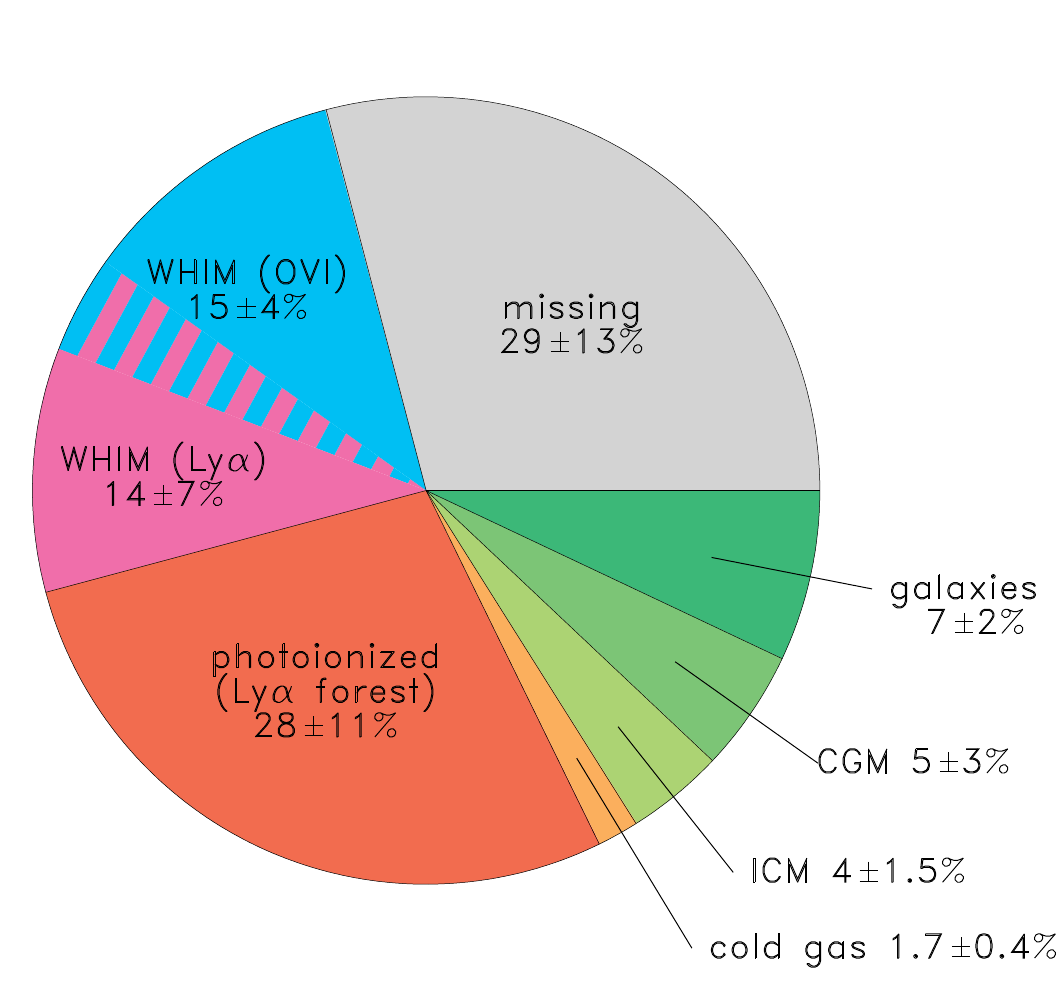}
   \caption{Compilation of current observational measurements of the low-redshift baryon census
   (Section 3.3).  Slices of the pie-chart show baryons in collapsed form (galaxies, groups, clusters), in 
   the circumgalactic medium (CGM) and intercluster medium (ICM) and in cold gas (\HI\ and \HeI). 
   Reservoirs include diffuse photoionized \Lya\ forest and WHIM traced by \OVI\ and broad \Lya\ 
   absorbers.  Blended colors (BLAs and \OVI) have combined total of $25\pm8$\%, accounting for 
   double-counting of WHIM at $10^5 - 10^6$~K with detectable metal ions.  The collapsed phases
   (galaxies, CGM, ICM, cold neutral gas) total $18\pm4$\%.  Formally, $29 \pm 13$\% of the baryons 
   remain unaccounted for.  Our simulations (Figure 2) suggest that an additional 15\%  reside in X-ray 
   absorbing gas at $T \geq 10^6$~K.   Additional baryons may be found in weaker lines of  low-column 
   density \OVI\ and \Lya\  absorbers.  Deeper spectroscopic UV and X-ray surveys are needed to 
   resolve this issue.  
     }
  \end{figure} 


What observations and theoretical work are needed to make progress on error bars or
baryon detections?   First, we need more precise UV absorption-line surveys to measure
\OVI\ and \Lya\ absorbers to lower column densities.  As described in Section~3, the numbers of
absorbers in current surveys become increasingly uncertain at column densities $\log N_{\rm HI} < 13.0$
and $\log N_{\rm OVI} < 13.5$.  The current surveys integrate below these levels, but our
experience using COS to re-examine earlier \Lya\ and \OVI\ detections with FUSE, GHRS, and
STIS, suggests that some of these weak absorbers are unconfirmed with high-S/N data.
In our \HST\ Archive Legacy Project, Tilton \etal\ (2012) reanalyzed \OVI\ and \Lya\ data from 
\HST/STIS  to provide critically evaluated column densities for $N_{\rm HI}$ and $N_{\rm OVI}$, 
and their absorption-line frequencies, $d {\cal{N}}/dz$.  The new \OVI\ data were included, together
with our fits to $f_{\rm OVI} \, (Z/Z_{\odot})$, to yield more accurate values of $\Omega_b^{{\rm (OVI)}}$.  

With the ten-fold increase in UV sensitivity throughput of Cosmic Origins Spectrograph on \HST,  
we should be able to do even better.  We hope to use  \HST/COS to obtain high-quality data 
(S/N $\geq 30$) to search for additional baryons in weak absorbers and constrain the predicted
flattening in the column density distributions of \HI\ and \OVI.    Scaled to column densities 
$(10^{13}~{\rm cm}^{-2}) N_{13}$, these lines have equivalent widths of (12.5~m\AA)$N_{13}$ for 
\OVI\ $\lambda1032$ and (54.5~m\AA)$N_{13}$ for \Lya.  These weak-absorber surveys will require
 \HST/COS sensitivity to 4~m\AA\ equivalent widths, which is achievable at S/N = 30 toward many 
 bright AGN background targets.     
We can also use COS to obtain better detections and statistics for broad \Lya\ absorbers (BLAs) and 
the \NeVIII\ doublet ($\lambda\lambda770.4, 780.3$).  The \NeVIII\ lines are potentially more
reliable probes of hot, collisionally ionized gas than \OVI, since \NeVIII\ requires 207~eV to produce
and is likely to be less contaminated by photoionization.  The lower solar neon abundance,
(Ne/O)$_{\odot} \approx 0.15$, makes the \NeVIII\  lines weak, and redshifts $z > 0.47$ are 
needed to shift them into the \HST/COS band.  The BLAs have considerable promise for
WHIM probes, as they do not require corrections for metallicity.  They do require determining
the neutral fraction, $f_{\rm HI}$, through careful modeling of the gas temperature and 
ionization conditions.

It would also be helpful to verify the claimed X-ray detections of \OVII\ in the WHIM (Nicastro
\etal\ 2005) which are not confirmed in other data (e.g., Kaastra \etal\ 2006; Rasmussen \etal\
2007) or in re-analysis of the same data (Yao \etal\ 2012).   The most critical observations for the
WHIM census may require a next generation of X-ray spectrographs to measure the weak absorption
lines of \OVII\ $\lambda21.602$, \OVIII\ $\lambda18.969$, and other He-like and H-like lines of 
abundant metals (\CV, \CVI, \NVI, \NVII).  As discussed by Yao \etal\ (2012), this requires high-throughput 
spectrographs ($E \approx 0.3-1.0$ keV) with energy resolution $E/\Delta E > 4000$ sufficient to 
resolve \OVII\  absorbers with m\AA\ equivalent width.  For weak lines, the predicted \OVII\
equivalent widths are $W_{\lambda} = (2.88~{\rm m\AA}) (N_{\rm OVII}/10^{15}~{\rm cm}^{-2})$.  
   
Finally, the IGM simulations can be improved in several aspects, in order to use them as even 
more reliable predictors of IGM parameters.  
In this paper, we computed individual (cell-by-cell) values of metallicity $Z$ and ionization
fraction $f_{\rm OVI}$, together with their statistical variations and co-variance.  Our simulations 
(Smith \etal\ 2011) were performed with somewhat larger box sizes ($50h^{-1}$~Mpc) and on a 
larger $1024^3$ grid, compared with previous studies of low-$z$ IGM thermodynamics.  Our \OVI\
conclusions appear to be robust, as gauged by convergence tests and runs with different methods 
of injecting feedback from star formation into the grid.  Post-processing these simulations allowed us to
provide more accurate corrections for the product, $(Z / Z_{\odot}) f_{\rm OVI}$, as a function
of column density, \NOVI.   These statistics can be performed for other key ions (\OVII, \NeVIII).  
We will extend our simulations through higher resolution, addition of discrete ionizing sources,
and radiative transfer.  We also will explore the the correct mixture of collisional ionization and 
photoionization in the WHIM, a project that requires understanding the implications of different
feedback mechanisms for injecting mass, thermal energy, and metals into the CGM.
How these metals mix and radiate likely determines the thermodynamics of the surrounding IGM.   
Because there still exist considerable differences between how simulations treat the thermal state of the
IGM, we will continue to push to higher resolution to capture the low-mass galaxies which are
important for metal production and mass injection.

\acknowledgments

This work was supported by NASA grant NNX08AC14G for COS data analysis and 
an STScI archival legacy grant AR-11773.01-A.    Our theoretical work and numerical
simulations were supported by the Astrophysical Theory Program (NNX07-AG77G  
from NASA and AST07-07474 from NSF) at the University of Colorado Boulder.
We thank Eric Hallman, Michele Trenti, John Stocke and Evan Tilton for comments on 
the manuscript, and Mark Voit  for discussions on hot gas in clusters and galaxy halos.  

\bigskip\bigskip

\section*{Appendix A:  \Lya\ Forest Baryon Content and Error Analysis}

Here, we summarize the current status of the error budgets for the major contributors to uncertainty
in the diffuse \Lya\ forest baryon census.  Many of the scalings enter the formula for $\Omega_b^{\rm (HI)}$ 
as the square root, a weak dependence arising because recombination theory predicts that 
$N_{\rm HI} \propto n_H^2$ for highly ionized absorbers.  Thus, the total baryon mass (proportional to $n_H$) 
depends on a weight factor $N_{14}^{1/2}$ (see equation [6]).  From this formulation, we can 
conduct an error-propagation analysis, to arrive at the overall uncertainty on the \Lya\ contribution to 
$\Omega_b$ in equation (9).   This quantity depends on five parameters, which we list below with assigned errors.  
\begin{enumerate}

\item {\bf Distribution of \HI\ column densities.}   Danforth \& Shull (2008) measured the distribution
of \HI\ column densities, $d{\cal {N}} / dz \propto N_{\rm HI}^{-\beta}$.  Based on 650 low-redshift \Lya\
absorbers, they found a slope $\beta = 1.73 \pm 0.04$ and a baryon content of $29\pm4$\% of
$\Omega_b$, integrated over $12.5 \leq \log N_{\rm HI} \leq 16.5$.   Our new survey (Tilton \etal\ 2012) 
finds $\beta = 1.68 \pm 0.03$ and a contribution of $28\pm4$\% from the \Lya-forest plus partial 
Lyman-limit systems to the baryon fraction.  We assume 15\% uncertainty in the column-density distribution 
statistics, which are increasingly uncertain at $\log N_{\rm HI} < 13.0$.   Extending the $N_{\rm HI}^{-1.7}$ 
distribution from $\log N = 12.5$ to 12.0 would add another 3\% to the baryon fraction (from 28\% to 31\%)
 if there is no change in slope. 
 
\item {\bf Hydrogen photoionization rate.}   This rate, $\Gamma_H$, depends on the radiation field 
normalization at 1 ryd ($I_0$) and spectral slope ($\alpha_s$) between 1.0--1.5 ryd.  Previous AGN
composite spectra found indices ranging from $\alpha_s = 1.76\pm0.12$ (Telfer \etal\ 2001) to 
$\alpha_s = 0.56^{+0.38}_{-0.28}$ (Scott \etal\ 2004).  New \HST/COS composite spectra of AGN find 
an index $\alpha_s = 1.59\pm0.20$ (Shull \etal\ 2012b), close to the Telfer \etal\ (2001) value for radio-quiet
AGN.   Our calculations also take into account the $(1+z)^{4.4}$ rise in $\Gamma_H(z)$ from $z = 0$ to 
$z = 0.7$, a fit to the recent calculations of Haardt \& Madau (2012).  We estimate the joint error on 
$\Gamma_H$ as $\pm 50$\% from models of the metagalactic radiation field from quasars and 
galaxies (Shull \etal\ 1999; Haardt \& Madau 2012).   

\item {\bf Characteristic scale-length of absorbers.}  The characteristic impact parameter, $p_{100}$, at
$N_{\rm HI} \approx 10^{14}$~\cd\ in units of 100 kpc, is inferred by direct and indirect means,  including 
comparing ``hits and misses" of \Lya\ absorbers along nearby sight lines and  the cumulative distributions 
of absorbers with nearest-neighbor galaxies (Stocke \etal\ 1995; Shull \etal\ 1998).    The frequency of  \Lya\ 
absorption lines per unit redshift also implies 200--300 kpc absorber cross sections, when associated with the space 
density of galaxies down to luminosities $L \approx 0.01-0.03L^*$ (Shull \etal\ 1996; Stocke \etal\ 2006; 
Prochaska \etal\ 2011).  We adopt an uncertainty of $\pm50\%$ on $p_{100}$.  

\item {\bf Electron temperature.}    Temperature ($T_e$) enters through the square root of the hydrogen 
recombination rate coefficient, $\alpha_H^{(A)} \propto T^{-0.726}$.   Models of IGM photoelectric 
heating (Donahue \& Shull 1991) predict a range from 5000~K to 30,000~K.  For $z < 0.4$, with small 
expected variations from photoionization, we adopt  an uncertainty of $\pm30$\%. 

\item {\bf Hubble constant.}   This parameter has been measured as $H_0 = 72\pm8$ km~s$^{-1}$~Mpc$^{-1}$ 
(Freedman \etal\ 2001) and $H_0 = 73.8\pm2.4$ km~s$^{-1}$~Mpc$^{-1}$ (Riess \etal\ 2011), using distance 
scales based on Cepheids in galaxies with Type Ia supernovae.  To be conservative, we adopt an 
error of $\pm5$\%.  

\end{enumerate}

From standard error-propagation formulae, we write the relative error on $\Omega_b^{(\rm HI)}$ 
as the quadrature sum of relative errors on the five parameters, weighted by the square of the exponents
(1, 0.5, or 0.363) as they appear in the scaling (see equation 9):
\begin{eqnarray}
    \left( \frac {\sigma_{\Omega}} {\Omega} \right) ^2 &=&  \left( \frac {\sigma_N}{N} \right)^2 
         +   \left( \frac {\sigma_h}{h} \right)^2 +  \left( \frac {\sigma_N}{N} \right)^2  +  (0.5)^2   \nonumber \\
        & \times &  \left[  \left( \frac {\sigma_{\Gamma}} {\Gamma} \right)^2 
         +   \left( \frac {\sigma_p}{p} \right)^2  \right]
         + (0.363)^2 \left( \frac {\sigma_T}{T} \right)^2 
\end{eqnarray}
This formula gives a relative error $(\sigma_{\Omega} / \Omega) = 0.40$, so that we can
express $\Omega_b^{(\rm HI)} / \Omega_b^{(\rm tot)} = 0.28 \pm 0.11$.   Most (77\%) of the error budget
comes from uncertainties in the ionizing radiation field ($\Gamma_H$) and characteristic absorber size 
($p_{100}$).

\newpage

\section*{Appendix B:   Current Baryon Census} 

\noindent
The following paragraphs summarize our current knowledge of the baryon-content in various components 
and thermal phases of the IGM.  

 \begin{itemize}
 
 \item {\bf Photoionized \Lya\ Absorbers.}   
 For this paper, we adopt $\Omega_b^{\rm (HI)} = 28 \pm 11$\% based on the results of Danforth \& 
Shull (2008), our new survey (Tilton \etal\ 2012), and the systematic uncertainties discussed in 
Section 3.2.  The mid-range distribution in column densities, $13.0 <  \log N_{\rm HI} < 14.5$, is fairly 
well characterized, but the numbers of high-column absorbers are small.  Their contribution to 
$\Omega_b$ remains uncertain owing to corrections for their size and neutral fraction.  At the
low end of the column-density distribution, there could be modest contributions to the baryon content 
from weaker \Lya\ absorbers.  In current surveys, their numbers are increasingly uncertain at 
$\log N_{\rm HI} < 13$ (our integration was down to 12.5).    
For a power-law distribution with $\beta = 1.7$, extending the distribution from $\log N = 12.5$ 
down to 12.0 would increase the baryon fraction by another 10\% (from 28\% to 31\%).  
For this paper (Figure 9), we adopt $\Omega_b^{\rm (HI)} = 28 \pm 11$\% based on the results of 
Danforth \& Shull (2008), Tilton \etal\ (2012), and the systematic uncertainties discussed in Section 3.2.  
    
 \item {\bf WHIM (\OVI-traced).}   
 Previous \FUSE\ surveys of intergalactic \OVI\ absorbers (Danforth \& Shull 2005; Tripp \etal\ 2006)
 found lower limits of  5\% and 7\%, respectively for the contribution of this gas to the baryon inventory.  
 These surveys assumed metallicity $Z = 0.1 Z_{\odot}$ and ionization fraction $f_{\rm OVI} = 0.2$.  
 In 2008, three \OVI\ surveys with \HST/STIS (Danforth \& Shull 2008; Tripp \etal\ 2008; Thom \& Chen 2008) 
 probed to lower \OVI\ column densities.  Based on the survey of 83 \OVI\ absorbers (Danforth \& Shull 2008) 
 and our more recent analysis of 111 \OVI\ absorbers (Tilton \etal\ 2012), we adopt a baryon fraction
 $\Omega_b^{(\rm OVI)} = 17 \pm 4$\%.  The factor-of-two increase arises primarily from our revised corrections, 
 $(Z / Z_{\odot}) \, f_{\rm OVI} = 0.01$,  for metallicity and \OVI\ ionization fraction (see Section 3.1).   There may 
 be some, as yet undetermined, overlap of photoionized \OVI\ with the \Lya\ forest.
 
\item {\bf WHIM (BLA-traced).}  
Broad \Lya\ absorbers (BLAs) were proposed (Richter \etal\ 2004, 2006;  Lehner \etal\ 2006, 2007) 
as repositories of a substantial fraction of the low-redshift baryons.  BLAs are defined as \Lya\ absorbers 
with Doppler parameters $b \geq 40$~\kms, corresponding to temperatures 
$T = (m_H b^2/2k) = (9.69 \times 10^4~{\rm K})b_{40}^2$ for pure thermal broadening 
with $b = (40$~\kms)$b_{40}$.  Owing to the large abundance of hydrogen,
a small neutral fraction (\HI) remains detectable in \Lya\ up to $T \sim10^6$~K, although 
high-S/N is required to measure the broad, shallow absorption.  As a result, the surveys of BLAs
differ considerably.  In their survey of seven AGN sight lines, Lehner \etal\ (2007) found a BLA 
frequency of $d{\cal N}/dz = 30 \pm 4$ for absorbers with 40~\kms\ $\leq b \leq 150$~\kms\ and  
$\log N_{\rm HI} \geq 13.2$.   They claimed that 20\% of the baryons reside in BLAs.  A more recent
survey (Danforth \etal\ 2010) came to different conclusions.   Surveying BLA candidates along seven 
AGN sight lines observed by HST/STIS, their BLA absorption-line frequency per unit redshift was 
$d{\cal N}/dz = 18\pm11$, comparable to that of the \OVI\ absorbers but 40\% lower than that found
by Lehnert \etal\ (2007).   After correction for possible (20--40\%) overlap between BLA and \OVI\ 
(metal-bearing) absorbers, the corresponding baryon fraction is
$\Omega_{\rm BLA} / \Omega_b  =  0.14^{+0.024}_{-0.018}$.  For Figure 9, we adopt a value of
$14\pm7$\%, with an increased error reflecting the uncertain detection statistics.  We apply their 
metallicity-based correction to obtain a blended total  (approximately 1/4 overlap) of $25\pm8$\%  
for the \OVI/BLA-traced WHIM over the temperature range $5 \leq \log T \leq 6$.  
 
 \item {\bf WHIM (X-ray absorber-traced).}  
 Our simulations (Figure 2) and those of Cen (2012) suggest that $\sim15$\% of the baryons could be 
 contained in hotter WHIM, at  $T > 10^6$~K.  Some of this gas could be detectable in 
 X-ray absorption lines of trace metals (\OVII, \OVIII, \NeIX,\ NeX, etc.).  However, most of these 
 weak absorption lines are below the detection limits of current X-ray spectrographs (Yao \etal\ 2012).   
 The suggested absorption systems are too few in number to provide good statistics, 
 and many are unconfirmed.  

 \item {\bf Galaxies.}  Salucci \& Persic (1999) found that galaxies contribute 7\% of the baryons. 
 More recent discussion by Fukugita \& Peebles (2004) estimated 6\%.   In Figure 9, we assume that 
 galaxies contribute $7\pm2$\% of the baryons.  
 
 \item {\bf Groups and Clusters.}   The integrated cluster mass function of Bahcall \& Cen (1993) was
 used by Fukugita \etal\ (1998) to find that the baryon contribution of clusters of galaxies consists
 of $\Omega_b^{\rm (stars)} = 0.00155h^{-1.5}$ and $\Omega_b^{\rm (gas)} = 0.003h^{-1}$.  Adjusting for
 $h = 0.7$, we find a cluster contribution of $\Omega_b^{\rm (cl)} = 0.00308$ or 6.8\% of the baryons.  
 Fukugita \& Peebles (2004) revised  the hot-baryon contribution to 
 $\Omega_b^{\rm (cl)} = 0.0018 \pm 0.0007$ or 4\% of the baryons, owing to a redefinition of 
 cluster mass (Reiprich \& B\"ohringer 2002).   For the total contribution of galaxy clusters, including
 their hot  gas, we adopt a fraction $4.0\pm1.5$\% of the baryons.  

 \item {\bf Cold H~I (and He~I) Gas.}   Zwaan \etal\ (2003) and Rosenberg \& Schneider (2003) conducted 
 blind \HI\ surveys in the 21-cm line that probe the mass density in neutral atomic gas.    The HIPASS 
 survey (Zwaan \etal\ 2003) found $\Omega_{\rm HI} = (4.7 \pm 0.7) \times10^{-4}$.  Following the 
 discussion of Fukugita \& Peebles (2004), which augments the \HI\ measurements by the expected
 accompanying cold \HeI\  and H$_2$, we plot the total cold  gas mass in Figure 9 as $1.7 \pm 0.4$\% 
 of the baryons.  
 
 \item {\bf Circumgalactic Medium (CGM). }   X-ray spectra of AGN taken with both {\it Chandra} and
 {\it XMM/Newton} detected strong \OVII\ absorbers at $z \approx 0$ (see Bregman 2007;  McKernan \etal\ 2005; 
 Wang \etal\ 2005; Fang \etal\ 2006).  The typical detected column densities,  $N_{\rm OVII} \approx 10^{16}$ \cd, 
 correspond to ionized hydrogen column densities $N_{\rm HII} \approx 10^{20}$ \cd\ assuming a length-scale 
 $\sim10$~kpc and mean metallicity of 20\% solar.   Several arguments suggest that the Galactic \OVII\ resides 
 in a thick disk or low scale-length halo (5--10 kpc;  see Yao \& Wang 2005).  Such a reservoir holds 
 $\sim10^9~M_{\odot}$ which is $\sim2$\% of the $5\times10^{11}~M_{\odot}$ in Milky Way baryons.   
 Bregman (2007) suggested that a hot gaseous medium with mass $\sim10^{10}~M_{\odot}$ might extend 
 throughout the Local Group.  Such a reservoir is also $\sim$2\% of the $6\times10^{11}~M_{\odot}$ baryonic 
 mass of the Local Group, assuming total mass $5\times10^{12}~M_{\odot}$ and 12\% baryon fraction 
 (McGaugh \etal\ 2010).   However, the Milky  Way halo and Local Group may not be typical of other 
 star-forming galaxies.   Recent COS studies of 42 galaxies (Tumlinson \etal\ 2011) found that large 
 (150 kpc) oxygen-rich halos of star-forming galaxies are major reservoirs of galactic metals.  
 Additional CGM could extend to distances of 100-200 kpc from galaxies with higher specific star formation 
 rates (Stocke \etal\ 2006).  Savage \etal\ (2010) detected hot circumgalactic gas ($\log T = 5.8-6.2$) in \HI-free 
 \OVI\ absorption associated with a pair (perhaps small group) of galaxies at $z \approx 0.167$, at impact parameters 
 $p \approx 100$~kpc.  They note that the absence of \HI\ absorption with broad  \OVI\ suggests a ``rare but important 
 class of low-$z$ intergalactic medium absorbers". 
 
 Prochaska \etal\ (2011) took this idea further, suggesting that {\it all} the \OVI\ arises in the  ``extended CGM 
 of sub-$L^*$ galaxies".   Their mass estimate associates the \OVI\ absorption with \HI\ around
 galaxies ($0.1 < L/L^* < L*$) 
 having constant hydrogen column densities, $N_H = 10^{19}$~\cd, over the full 200--300 kpc extent,
 as gauged by AGN-galaxy impact parameters.   They estimated that these extended halos could contain a mass 
$M_{\rm CGM} \approx (3 \times10^{10}~M_{\odot})(r_{\rm CGM}/300~{\rm kpc})^2$, which would represent
$\sim50$\% of the baryon masses of  present-day sub-$L^*$ galaxies.  They further assert that gas at these 
large distances is gravitationally bound and virialized.  We question the realism of constant-$N$, virialized 
gas at such large distances from dwarf galaxies.  Previous nearest-neighbor studies low-$z$ \OVI\ absorbers and 
 galaxies (Stocke \etal\ 2006; Wakker \& Savage 2009) found correlations with $L^*$ galaxies (at 800 kpc) 
 and with dwarf ($0.1L^*$) galaxies (at 200 kpc).   For Figure 9, we adopt a CGM contribution of $5\pm3$\%,
 recognizing that the CGM reservoir is still poorly understood.  
   
 \end{itemize}


\end{document}